\documentclass[reprint,aps,prc,twocolumn,superscriptaddress,floatfix,10pt]{revtex4-2}

\usepackage{bm}
\usepackage{dcolumn}
\usepackage{amsthm}
\usepackage{amssymb}
\usepackage{amsmath}
\usepackage{graphicx}
\usepackage{xcolor}
\usepackage{color}
\usepackage{fix-cm}
\usepackage{mathptmx} 
\usepackage[T1]{fontenc}
\usepackage[colorlinks,allcolors=blue]{hyperref}
\usepackage[utf8]{inputenc}

\usepackage{CJK}

\makeatletter
\def\NAT@def@citea{\def\@citea{\NAT@separator}}
\makeatother

\begin{document}
\begin{CJK*}{UTF8}{}
\title{Microscopic study of the compound nucleus formation in cold-fusion reactions}

\author{Xiang-Xiang Sun~\CJKfamily{gbsn}(孙向向)}
\affiliation{School of Nuclear Science and Technology,
 University of Chinese Academy of Sciences,
 Beijing 100049, China}
 \affiliation{CAS Key Laboratory of Theoretical Physics,
              Institute of Theoretical Physics,
              Chinese Academy of Sciences,
              Beijing 100190, China}
\author{Lu Guo~\CJKfamily{gbsn}(郭璐)}
\email{luguo@ucas.ac.cn}
\affiliation{School of Nuclear Science and Technology,
University of Chinese Academy of Sciences,
Beijing 100049, China}
\affiliation{CAS Key Laboratory of Theoretical Physics,
              Institute of Theoretical Physics,
              Chinese Academy of Sciences,
              Beijing 100190, China}

\begin{abstract}
The understanding of the fusion probability is of particular
importance to reveal the mechanism of producing superheavy elements.
We present a microscopic study of the compound nucleus formation
by combining time-dependent density functional theory,
coupled-channels approach, and dynamical diffusion models.
The fusion probability and
compound nucleus formation cross sections
for cold-fusion reactions $^{48}$Ca+$^{208}$Pb,
$^{50}$Ti+$^{208}$Pb, and $^{54}$Cr+$^{208}$Pb
are investigated
and it is found that the deduced capture barriers,
capture cross sections for these reactions
are consistent with experimental data.
Above the capture barrier, our calculations reproduce the measured
fusion probability reasonably well.
Our studies demonstrate that the restrictions from
the microscopic dynamic theory
improve the predictive power of
the coupled-channels and diffusion calculations.
\end{abstract}
\maketitle
\end{CJK*}

\section{Introduction}
The study on the superheavy elements (SHEs) is one of the most
important topics in nuclear physics nowadays
\cite{Hofmann2000_RMP72-733,Hamilton2013_ARNPS63-383,
Oganessian2015_RPP78-036301,Giuliani2019_RMP91-011001}.
The cross section of
fusion-evaporation reactions
for producing superheavy nuclei (SHN)
is extremely small, in the order of $10^{-36}$ cm$^2$
and strongly dependent on the combination of two colliding
nuclei and the incident energy.
Up to now, SHEs up to $Z=118$ have been synthesized by employing
two types of fusion reactions in the laboratory:
the cold-fusion reactions with $^{208}$Pb and $^{209}$Bi as targets
\cite{Hofmann2000_RMP72-733,Morita2004_JPSJ73-2593}
and hot-fusion ones between $^{48}$Ca and actinide nuclei
\cite{Oganessian2007_JPG34-R165,Oganessian2010_PRL104-142502}.
The former leads to the discoveries of new elements up to $Z=113$
and the later for SHN with $113\le Z \leq 118$ so far.
Lots of efforts have been made to produce SHN with $Z=119$
and $Z=120$ \cite{Oganessian2009_PRC79-024603,
Kozulin2010_PLB686-227,
Hofmann2016_EPJA52-180,
Khuyagbaatar2020_PRC102-064602,
Albers2020_PLB808-135626},
but there is no evidence for the synthesis of the new elements.

It is significant to understand the reaction dynamics to choose the appropriate
combination of projectile and target nuclei to produce new SHN.
Conceptually, the fusion-evaporation reaction can be
divided into three steps,
namely the capture process, the fusion process,
and the de-excitation of the excited compound nucleus (CN) against fission
and light particle emission.
The capture process is generally treated as
a one-dimensional quantum tunneling under a given ion-ion potential
with considering the couplings to the excitations of
projectile and target nuclei
\cite{Hagino2012_PTP128-1061,Back2014_RMP86-317360}
and the results from semi-empirical systematic
\cite{Swiatecki2005_PRC71-014602,
Zagrebaev2015_NPA944-257,
Wang2017_ADNDT114-281370}
and coupled-channels calculations
\cite{Hagino2012_PTP128-1061}
are well consistent with measurements
\cite{Clerc1984_NPA419-571,Pacheco1992_PRC45-28612,
Prokhorova2008_NPA802-45,Banerjee2019_PRL122-232503}.
For the de-excitation process,
the survival probability of CN
is well studied with the statistical models
\cite{Adamian2000_PRC62-064303,Zagrebaev2015_NPA944-257}
and the dependence on reaction parameter is also well understood,
although it is strongly influenced by the fission barrier height.
Therefore, most of theoretical approaches
used to estimate the evaporation-residue cross sections have similar
conclusions on these two steps
\cite{Antonenko1993_PLB319-425,Adamian1997_NPA627-361,
Swiatecki2005_PRC71-014602,
Feng2006_NPA771-50,Zhu2014_PRC89-024615,
Wang2012_PRC85-041601R,Liu2011_PRC84-031602R,
Cap2011_PRC83-054602,Cap2013_PRC88-037603,
Zagrebaev2015_NPA944-257,
Bao2017_PRC96-024610,Hagino2018_PRC98-014607,
Lv2021_PRC103-064616}.

However, the fusion process is still not well understood up to now.
The calculated fusion probabilities with empirical formulae,
master equations, or diffusion models
differ by several orders of magnitude
\cite{Adamian2000_PRC62-064303,Swiatecki2005_PRC71-014602,
Feng2006_NPA771-50,
Loveland2007_PRC76-014612,
Naik2007_PRC76-054604,
Wang2012_PRC85-041601R,
Yanez2013_PRC88-014606,
Lu2016_PRC94-034616,
Lv2021_PRC103-064616}.
Even more, the dependence of the fusion probability on
the reaction entrance channel is not well established
\cite{Naik2007_PRC76-054604,Zagrebaev2008_PRC78-034610,
Yanez2013_PRC88-014606}.
Experimentally, the fusion probability can be extracted from the measurement
of fusion-evaporation residue cross sections
\cite{Berriman2001_Nature413-144,Khuyagbaatar2012_PRC86-064602},
comparing the
width of fragment mass distribution with the width expected
in the case of pure fusion-fission
\cite{Lin2012_PRC85-014611,Hammerton2015_PRC91-041602,Banerjee2019_PRL122-232503},
or the analysis of the
fragment angular distribution
\cite{Back1985_PRC31-2104,Tsang1983_PLB129-18,Banerjee2021_PLB820-136601}.
Recent years, many efforts have been made
to measure the fusion probability
and lots of progresses have been achieved \cite{Hinde2021_PPNP118-103856}.
Very recently, experimentalists have extracted
the fusion probabilities
for both cold-fusion and hot-fusion
reactions \cite{Naik2007_PRC76-054604,Banerjee2019_PRL122-232503,
Banerjee2021_PLB820-136601}.
These measurements provide new constraints
to theoretical investigations for the synthesis of
SHN.

It has been shown that the macroscopic models can reproduce
the cross section data, but a lot of adjustable parameters are involved.
In addition, the lack of dynamical effect challenges
its predictive power for
reactions without available experimental data.
Modern dynamical microscopic approaches can provide
insight into the low-energy heavy-ion collisions
\cite{Umar2006_PRC74-021601R,Wen2013_PRL111-12501,
Simenel2013_PRC88-064604,Washiyama2008_PRC78-024610}.
The time-dependent Hartree-Fock (TDHF) approach
has been successfully applied to study many aspects of
low-energy heavy-ion collisions
(see Refs. \cite{Simenel2012_EPJA48-152,
Nakatsukasa2016_RMP88-45004,
Simenel2018_PPNP103-19,
Stevenson2019_PPNP104-142164,
Sekizawa2019_FP7-20} and references therein).
Based on a mass of TDHF simulations,
the fusion probability has been estimated by using
the sharp cut-off approximation for
$^{48}$Ca+$^{239,244}$Pu at several selected
incident energies \cite{Guo2018_PRC98-064609}.
However, this method is restricted
for a systematic study
on fusion probabilities
due to its computational cost.
It has been shown that
the TDHF simulations can provide the main
ingredients of coupled-channels calculations for capture cross section
\cite{Simenel2013_PRC88-024617,Guo2018_PLB782-401}
and diffusion processes
\cite{Sekizawa2019_PRC99-051602R}.
Note that the TDHF approach is not capable for the
whole process of fusion-evaporation reactions.
The purpose of the present work is to
combine TDHF with both the coupled-channels and
dynamic diffusion approaches, aiming the study of fusion
probability systematically.
In this approach, one can eliminate the uncertainties
of adjustable parameters for calculating capture and fusion cross sections
under the restriction from microscopic TDHF theory, meanwhile
the influences of the structures of reactants and dynamical effects
can be taken into account.
In this work, we use this method to study the systematics of the fusion
probability of cold-fusion reactions.
This article is organized as follows.
In Sec.~\ref{Sec2},
we show the main theoretical formulation to calculate capture cross sections and
fusion probabilities.
Section~\ref{Sec3} presents the calculational details and the discussion of results.
A summary and perspective is provided in Sec.~\ref{Sec4}.

\section{Theoretical framework}
\label{Sec2}

In the TDHF theory,
the Hamiltonian $\hat{H}$ is a functional of densities
and the dynamic process is described by the evolution of the
one-body density $\hat{\rho}$, which is obtained by solving the TDHF equation
\begin{equation}
  i\hbar \frac{\partial }{\partial t} \hat\rho = \left[\hat{H}[\hat{\rho}],\hat{\rho}\right].
\end{equation}
Since the TDHF theory describes the collective motion in a
semiclassical way,
the quantum tunneling of the many-body wave function is not included.
Therefore when studying capture cross sections by using the TDHF theory,
a commonly applied and very effective strategy is
using the ion-ion potential obtained from
frozen HF \cite{Guo2012_EWC38-09003},
density constrained (DC) TDHF \cite{Umar2006_PRC74-021601R,
Umar2008_PRC77-064605,
Umar2010_PRC81-064607,
Simenel2013_PRC88-064604,
Simenel2013_PRC88-024617,
Umar2014_PRC89-034611,
Umar2016_PRC94-024605,
Guo2018_PLB782-401,Guo2018_PRC98-064607,
Godbey2019_PRC100-054612},
DC frozen HF \cite{Simenel2017_PRC95-031601R},
dissipative-dynamics TDHF \cite{Washiyama2008_PRC78-024610},
the Thomas-Fermi approximation
\cite{Wang2006_PRC74-044604},
and the Woods-Saxon (WS) potential with the fitted parameters
\cite{Simenel2013_PRC88-024617}
as the input of the coupled-channels code CCFULL
\cite{Hagino1999_CPC123-143}
to calculate the penetration probability.
In our approach, for a given colliding system,
we perform TDHF calculations to determine
the capture barrier, which is minimum incident energy
that the projectile can be captured by the target in TDHF simulation.
Then the calculated densities in their ground states and
the obtained capture barriers are used to fix the parameters
of the WS potential
\cite{Wong1973_PRL31-766}
\begin{equation}
V(R) = -\frac{V_0}{1+\exp\frac{R-
r_{0\mathrm{P}}A_\mathrm{P}^{1/3}-
r_{0\mathrm{T}}A_\mathrm{T}^{1/3}}{a}},
\end{equation}
with the depth $V_0$,
the diffuseness parameter $a$,
radius parameter for target (projectile)
$r_{0\mathrm{T}}$ ($r_{0\mathrm{P}}$),
and the mass number $A_\mathrm{T}$ ($A_\mathrm{P}$) of target (projectile).
In this work, the diffuseness parameter is determined by
\begin{equation}
  a=\frac{1}{1.17
  \left[ 1+0.53
  \left(A_\mathrm{P}^{-1/3}+A_\mathrm{T}^{-1/3}
  \right)
  \right]}\ \mathrm{fm},
\end{equation}
taken from Ref. \cite{Wang2017_ADNDT114-281370}
and the determination of $V_0$, $r_{0\mathrm{T}}$,
and $r_{0\mathrm{P}}$ will be introduced later.
Considering the coupling to low-lying states of
projectile and target nuclei,
the capture cross sections are obtained by using the
standard coupled-channels calculations
\cite{Hagino1999_CPC123-143}
and read
\begin{equation}
  \sigma_\mathrm{cap}(E_\mathrm{c.m.}) =
  \frac{\pi}{k^2}\sum_{J}
  (2J+1)T_J(E_\mathrm{c.m.}),
\end{equation}
where $k=\sqrt{2\mu E_\mathrm{c.m.}/\hbar^2}$ with $\mu$ being
the reduced mass in the entrance channel. $T_J(E_\mathrm{c.m.})$,
which is calculated by using the incoming wave boundary condition method
\cite{Hagino1999_CPC123-143},
is the penetration probability for given
incident energy $E_\mathrm{c.m.}$ and angular momentum $J$.
It should be noted that one can also use
the DC-TDHF method to extract the microscopic internuclear potentials,
but which needs much computational cost \cite{Umar2016_PRC94-024605}
for the reactions to produce SHN
and the obtained potential depends on the incident energy.
For a systematic study of fusion cross sections of cold-fusion reactions,
the computational cost of the DC-TDHF is too high.

For the second stage of the fusion-evaporation reactions,
the contact configuration of touching nuclei
can be transformed to CN configuration by overcoming the
inner barrier.
In this process, before the formation of CN,
quaifission happens and
the heavy nuclear system is split into two fragments.
Therefore,
the formation of CN is strongly hindered by quasifission,
which is dependent on the structure of the reactants and
has also been studied by using the TDHF approach
\cite{Oberacker2014_PRC90-054605,
Wakhle2014_PRL113-182502,Umar2015_PRC92-024621,
Umar2016_PRC94-024605,
Sekizawa2016_PRC93-054616,Guo2018_PRC98-064609}.
In the
fusion-by-diffusion (FbD) model
\cite{Swiatecki2003_APPB34-2049,
Swiatecki2005_PRC71-014602,
Cap2011_PRC83-054602,
SiwekWilczynska2012_PRC86-014611},
the formation of CN is idealized as a one-dimensional
diffusion over a parabolic barrier and the corresponding
probability is totally determined by the distance between
the surfaces of two colliding nuclei
at the injection point.
This quantity can be estimated from the TDHF evolution by choosing
the moment when the collective kinetic energy is completely
dissipated into the internal degrees of freedom in the over-damped
regime for a given incident energy,
see Ref. \cite{Sekizawa2019_PRC99-051602R} for details.
It should be mentioned that in Ref. \cite{Sekizawa2019_PRC99-051602R}
only $s$-wave scattering are considered and the penetration possibilities is obtained
for one or two selected incident energies at above-barrier region.
In our approach, the fusion probability $P_\mathrm{CN}(E_\mathrm{c.m.},J)$
at each angular momentum and incident energy
is obtained by using the formulas given in Ref. \cite{Cap2011_PRC83-054602}.

The CN formation cross section is
\begin{equation}
  \sigma_\mathrm{fus}(E_\mathrm{c.m.}) =
  \frac{\pi}{k^2}\sum_{J}
  (2J+1)T_J(E_\mathrm{c.m.})
  P_\mathrm{CN}(E_\mathrm{c.m.},J).
\label{eq:fus}
\end{equation}
Thus, one can study the systematics of the fusion
probability by comparing the effective fusion probabilities $P_\mathrm{fus}$ defined as
$\sigma_\mathrm{fus}/\sigma_\mathrm{cap}$ with measured ones.

\section{Results and discussions}
\label{Sec3}
In a recent experiment \cite{Banerjee2019_PRL122-232503},
the upper limits of
fusion probabilities
for cold-fusion
reactions of $^{48}$Ca, $^{50}$Ti, and $^{54}$Cr with
$^{208}$Pb have been extracted
systematically.
Therefore it is quiet interesting to examine whether
our approach is valid for these reactions.
The time evolution of each reaction system is
obtained by using the modified version of
the Sky3D code
\cite{Maruhn2014_CPC185-2195},
which was also used to perform calculations in
Refs. \cite{Dai2014_PRC90-044609,Guo2018_PLB782-401,
Guo2018_PRC98-064609,Guo2018_PRC98-064607,
Wu2019_PRC100-014612,Godbey2019_PRC100-054612,Li2019_SCPMA62-122011,
Wu2020_SCPMA63-242021,Wu2022_PLB825-136886}.
Recently, the density functional SLy5
\cite{Chabanat1998_NPA635-231} has been adopted
in many investigations
\cite{Umar2006_PRC74-021601R,Guo2018_PRC98-064609,Guo2018_PRC98-064607,
Guo2018_PLB782-401,Godbey2019_PRC100-054612,
Wu2020_SCPMA63-242021,Wu2022_PLB825-136886,
Sun2022_PRC105-034601} and also used in present work.
The ground states of $^{48}$Ca,
$^{50}$Ti,
$^{54}$Cr, and $^{208}$Pb are obtained by solving
the static HF equation on a three-dimensional grid
$28\times 28 \times 28$ fm$^3$.
To get a spherical ground state,
the proton orbital $1f_{7/2}$ is fully filled
with the equal weight for $^{50}$Ti and
the proton orbital $1f_{7/2}$ and
neutron orbital $2p_{3/2}$ are equally occupied
for $^{54}$Cr.
We mention that $^{50}$Ti is spherical due to the magic number $N=28$
and $^{54}$Cr has a prolate shape with quadrupole
deformation parameter of 0.21 in static HF calculations.
Thus this filling approximation is suitable for $^{50}$Ti.
Since in this work we disregard the
orientation effects of the projectile,
$^{54}$Cr is also set to be spherical by the same strategy used
in Ref. \cite{Sekizawa2019_PRC99-051602R}.
The dynamic evolution for these three
reactions are performed in a three-dimensional
grid with the size of $60 \times 40 \times 40$ fm$^{3}$.
The grid spacing in each direction is taken to be 1 fm
and the time step is 0.2 $\mathrm{fm}/c$.
All the numerical conditions have been checked for
achieving a good numerical accuracy for all the cases studied here.

\begin{figure}[htbp]
  \centering
  \includegraphics[width=.5\textwidth]{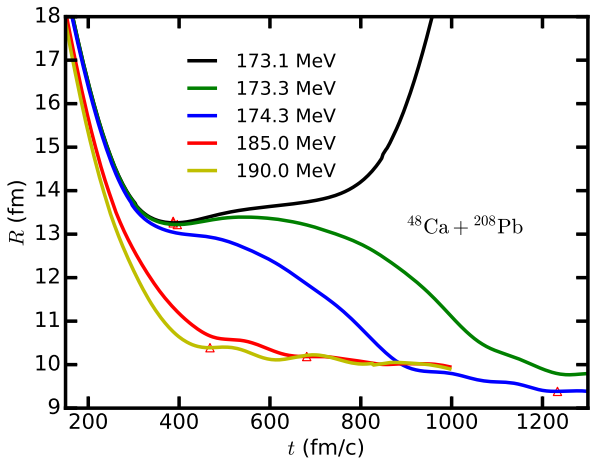}
  \caption{Time evolution of the distance between the fragments in $^{48}$Ca+$^{208}$Pb central collisions.
  The open triangles represent the first moment with a zero canonical momentum. }
\label{r-t}
\end{figure}

To get the Coulomb barrier of the each reaction system,
we perform TDHF simulations of central collisions
at different incident energies and around the barrier
the step of the incident energy is taken to be 0.2 MeV.
In Fig. \ref{r-t}, we show
the time evolution of the separation distance between the fragments for
$^{48}$Ca+$^{208}$Pb central collisions at
several selected incident energies.
This distance is determined by using the standard TDHF approach of finding left and
right dividing planes and is the separation distance
between the centers of the density in these two halves
\cite{Umar2006_PRC74-021601R,Maruhn2014_CPC185-2195}.
It should be mentioned that the time period 1300 fm$/c$ is enough
for these systems to come over the Coulomb barrier and to form the
contact configuration, i.e., the projectile captured by the target.
If one wants to get the fusion threshold energy or to judge quasifission, a
much longer time TDHF simulation is needed \cite{Washiyama2015_PRC91-064607}.
It is clear that when $E_\mathrm{c.m.} \geq 173.3$ MeV,
capture happens for
$^{48}$Ca+$^{208}$Pb,
in which the collective kinetic energy can be
entirely converted into the internal excitation of the contacting system.
Thus, the capture threshold energy $V_B^\mathrm{TDHF}$ for this reaction is 173.3 MeV.
The first moment corresponding to a zero conjugate momentum of $R$
at each reaction energy
is represented by the open triangles in Fig. \ref{r-t}.
We also perform similar calculations for $^{50}$Ti+$^{208}$Pb and $^{54}$Cr+$^{208}$Pb
and the corresponding threshold energies are 191.4 MeV and 209.4 MeV, respectively.
The Coulomb barrier for these three reactions obtained from such TDHF calculations
reasonably agree with those deduced from experimental capture cross sections
by using a classical barrier-passing model for fissionlike cross sections,
$173.4 \pm 0.1$, $192.6 \pm 0.1$, and
$207.3 \pm 0.3$ MeV for $^{48}$Ca, $^{50}$Ti, and $^{54}$Cr, respectively
\cite{Banerjee2019_PRL122-232503}.

\begin{figure*}[htbp]
  \centering
  \includegraphics[width=0.8\textwidth]{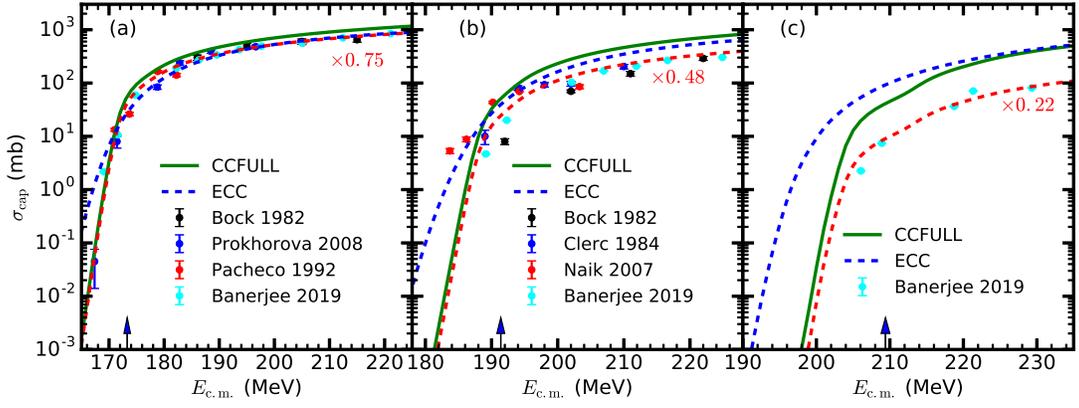}
  \caption{Calculated capture cross sections for
  $^{48}$Ca+$^{208}$Pb (a),
  $^{50}$Ti+$^{208}$Pb (b),
  and $^{54}$Cr+$^{208}$Pb (c).
  The calculated values with the ECC model
  \cite{Wang2017_ADNDT114-281370}
  and
  the experimental data taken from 
  Bock 1982: \cite{Bock1982_NPA388-334}, 
  Prokhorova 2008: \cite{Prokhorova2008_NPA802-45}, 
  Pacheco 1992: \cite{Pacheco1992_PRC45-28612}, 
  Clerc 1984: \cite{Clerc1984_NPA419-571},
  Naik 2007: \cite{Naik2007_PRC76-054604}, and
  Banerjee 2019: \cite{Banerjee2019_PRL122-232503}
  are shown for comparison.
  The capture thresholds are labelled by the up arrows for each reaction.
  The scaled cross sections are shown by the red dashed lines.
 }
\label{cap}
\end{figure*}

The radius parameters of the WS potential can be obtained from the ground-state density
in static HF calculations. Taking $^{208}$Pb as an example,
the radius from the center of nucleus to the isosurface
with half of the saturation density
($\rho_0=0.16\ \mathrm{fm}^{-3}$) is 6.672 fm.
Then its radius parameter
is $r_{0\mathrm{T}} = 6.67\times208^{-1/3}\ \mathrm{fm} = 1.126\ \mathrm{fm}$.
Similarly, we have $r_{0\mathrm{P}}=$
1.161 fm, 1.088 fm, and 1.087 fm
for $^{48}$Ca, $^{50}$Ti,
and $^{54}$Cr, respectively.
The depth $V_0$ can be adjusted for reproducing the capture thresholds
from TDHF calculations and they are 165.42 MeV, 251.998 MeV,
and 225.07 MeV for reactions $^{48}$Ca+$^{208}$Pb,
$^{50}$Ti+$^{208}$Pb, and $^{54}$Cr+$^{208}$Pb, respectively.
It should be noted that the low-lying
collective vibrations can also be estimated by the TDHF evolution with external fields.
But for the systems in question, the low-lying spectra of these nuclei are well known.
Therefore, we use the fitted WS potentials together with the experimental data of
excitation energies and deformations
taken to be the same as those provided in the supplement
of Ref. \cite{Banerjee2019_PRL122-232503},
to calculate
the capture cross section with the code CCFULL.
Finally, the calculated capture cross sections
$\sigma_\mathrm{cap}$ are shown in Fig. \ref{cap}
and compared with experimental data and the results from the ECC model
\cite{Wang2017_ADNDT114-281370}.
As we can see, for each reaction,
our calculations are in agreement with the ECC results
and reproduce the experimental data reasonably well.

\begin{figure*}[htb]
  \centering
  \includegraphics[width=0.8\textwidth]{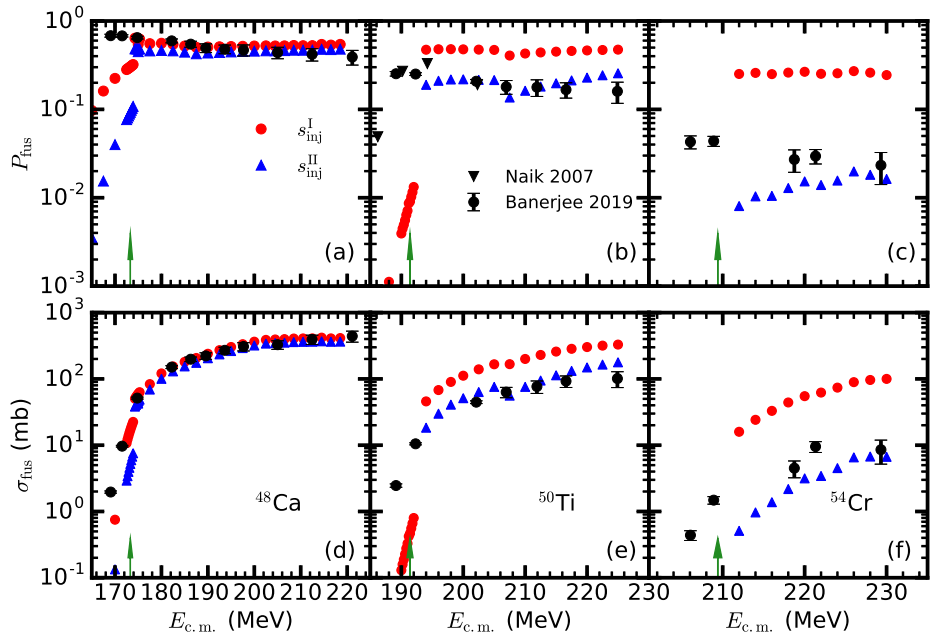}
  \caption{The effective fusion probabilities (the upper panel) and
  the fusion cross sections (the bottom panel) for
  $^{48}$Ca+$^{208}$Pb (a, d),
  $^{50}$Ti+$^{208}$Pb (b, e),
  and $^{54}$Cr+$^{208}$Pb (c, f).
  Red circles (blue triangles) represent
  the calculated results with $s_\mathrm{inj}^\mathrm{I}$
  ($s_\mathrm{inj}^\mathrm{II}$).
  The capture thresholds are labelled by the up arrows for each reaction.
  Experimental data taken from
  Banerjee 2019: \cite{Banerjee2019_PRL122-232503}
  are shown by solid points with error bars.
  For $^{50}$Ti+$^{208}$Pb, the measurements from Naik 2007: \cite{Naik2007_PRC76-054604}
  are presented by black triangles.
  }
\label{sig}
\end{figure*}

In the recent work \cite{Banerjee2019_PRL122-232503}, the fissionlike cross
sections are measured and also shown in Fig. \ref{cap}. The corresponding capture cross sections are obtained by using the CCFULL calculations with the capture barriers from
the classical barrier-passing model and experimental low-lying excitation information.
It is shown that the capture cross sections are larger than the measured fissionlike cross sections
and after scaling with suppression factors $S$, good agreements are achieved.
This suppression might be due to the energy dissipation before reaching the Coulomb barrier
\cite{Newton2004_PLB586-219,Newton2004_PRC70-024605}.
Therefore, we also multiply capture cross sections in our calculations by the same suppression factors used in Ref. \cite{Banerjee2019_PRL122-232503} and find scaled cross sections are more consistent with the measurements, shown in Fig. \ref{sig}.
In conclusion, by deducing the parameters involved in
coupled-channels calculations from TDHF simulations,
the capture cross sections can be well determined
for both below- and above-barrier regions. These results
demonstrate the effectiveness of our model.

\begin{figure}[htb]
  \centering
  \includegraphics[width=0.5\textwidth]{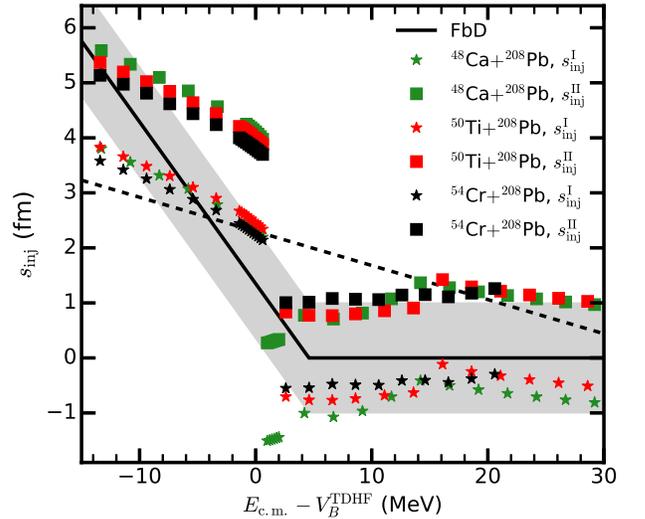}
  \caption{The injection parameters
  $s_\mathrm{inj}^\mathrm{I}$ (stars) and
  $s_\mathrm{inj}^\mathrm{II}$ (squares)
  for reactions $^{48}$Ca+$^{208}$Pb (green),
  $^{50}$Ti+$^{208}$Pb (red),
  and $^{54}$Cr+$^{208}$Pb (black) obtained from TDHF simulations.
  $s_\mathrm{inj}$ from the empirical formulae given in Refs. \cite{Cap2021_arXiv2107.00579}
  and \cite{Cap2011_PRC83-054602}
  are labelled by black solid and dashed lines, respectively.
  The shaded area represents an error
  corridor of $\pm 1$ fm for the formulae given in
  Ref. \cite{Cap2021_arXiv2107.00579} and is drawn to guide the eye.
  }
\label{sinj}
\end{figure}

For the calculations of $P_\mathrm{CN}$,
the only input parameter in the FbD model is the injection parameter,
which can be estimated by TDHF evolution \cite{Sekizawa2019_PRC99-051602R}
and is defined as
\begin{equation}
  s_\mathrm{inj} = R_\mathrm{min} - R_\mathrm{P} - R_\mathrm{T},
\end{equation}
where $R_\mathrm{min}$ is the distance between two fragments at the injection point.
In this work, we choose the first moment
with a zero canonical momentum in the TDHF evolution as the injection point.
In Fig. \ref{r-t}, we show the injection points
labelled by open triangles
for $^{48}$Ca+$^{208}$Pb for several selected incident energies.
In this way, $R_\mathrm{min}$ is determined.
The radii of colliding nuclei are estimated by using the empirical formula
$R_i = r_{0i}A_i^{1/3}\ (i=\mathrm{P},\mathrm{T})$
or taken to be the same as the root-mean-square matter radii in static HF calculation,
which are 3.56 fm, 3.59 fm, 3.67 fm, and
5.55 fm for $^{48}$Ca, $^{50}$Ti, $^{54}$Cr, and $^{208}$Pb, respectively.
The calculated injection parameters corresponding to the empirical formula of radius are labelled as
$s_\mathrm{inj}^\mathrm{I}$ and $s_\mathrm{inj}^\mathrm{II}$ for those with radii from static HF results.
The values of $s_\mathrm{inj}$ from the empirical formulae used in the FbD model
\cite{Cap2021_arXiv2107.00579,
Cap2011_PRC83-054602} are
also shown for comparison.
When $E_\mathrm{c.m.}-V^\mathrm{TDHF}_B>4.59$ MeV,
$s_\mathrm{inj}$ from the empirical formula
given in Ref. \cite{Cap2021_arXiv2107.00579}
is taken to be zero.
The shaded area in Fig. \ref{sinj} means a derivation of $\pm 1$ fm for the formula in Ref. \cite{Cap2021_arXiv2107.00579}.

In TDHF calculations,
the contact configuration can only be reached in above-barrier reactions
while the elastic scattering happens in below-barrier ones.
This leads to a sudden change of $s_\mathrm{inj}$.
It is found that for below-barrier region,
most values of $s_\mathrm{inj}^\mathrm{I}$
are located in the shaded area while $s_\mathrm{inj}^\mathrm{II}$
gives a much larger distance at the injection points
compared with the empirical formula in FbD model.
Above the barrier, the values of $s_\mathrm{inj}^\mathrm{I}$
are negative and most of them are well located at the shaded area,
indicating that the overlap of the densities for two colliding nuclei is
very large and resulting in a relatively large $P_\mathrm{CN}$.
For $s_\mathrm{inj}^\mathrm{II}$,
all of them are positive, indicating a small overlap of the densities and
smaller $P_\mathrm{CN}$ compared with these for $s_\mathrm{inj}^\mathrm{I}$.
Further more, $s_\mathrm{inj}$ extracted from TDHF calculations and the excess of
incident energy also hold a linear relation in below-barrier region,
which is consistent with that fitted to experimental data shown in FbD model
\cite{Swiatecki2005_PRC71-014602,Cap2011_PRC83-054602,Cap2013_PRC88-037603,
Cap2014_PLB736-478,Hagino2018_PRC98-014607,Cap2021_arXiv2107.00579}.
In above-barrier region, most values of $s_\mathrm{inj}^\mathrm{I}$ and $s_\mathrm{inj}^\mathrm{II}$ are
close to $-$0.6 fm and 1 fm, respectively, while
the closet distance is usually taken to be 0 in FdD calculations.
In our method, $s_\mathrm{inj}$ is no longer an adjustable parameter in FbD model, thus eliminating the uncertainties of the fusion cross section originated from $s_\mathrm{inj}$.
Generally, the energy dependence behavior of $s_\mathrm{inj}$ becomes weaker in above-barrier
collisions compared with those in below-barrier region,
i.e., the linear relation holding in below-barrier region disappears in above-barrier collisions.
This energy dependence behavior and the relevance with entrance channel
need to be explored further.

After obtaining the injection points,
by using formulas given in Ref.
\cite{Cap2011_PRC83-054602} with
the nuclear data tables
in Ref. \cite{Jachimowicz2021_ADNDT138-101393}
as inputs,
we calculate the fusion probabilities $P_\mathrm{CN}(E_\mathrm{c.m.},J)$
and fusion cross sections $\sigma_\mathrm{fus}$.
In order to compare with the upper limits of
measured fusion probabilities $P_\mathrm{sym}$ \cite{Banerjee2019_PRL122-232503},
we calculate the effective fusion probabilities $P_\mathrm{fus}$,
defined as the ratio between fusion cross sections and capture cross sections in our model,
which is independent on angular momenta.
The upper panel of
Fig. \ref{sig} shows the comparison between calculated $P_\mathrm{fus}$ and
$P_\mathrm{sym}$ taken from Ref. \cite{Banerjee2019_PRL122-232503}.
For $^{50}$Ti+$^{208}$Pb, the measurements given in Ref. \cite{Naik2007_PRC76-054604}
are presented by black triangles and shown in Fig. \ref{sig}(b).
The discontinuity of calculated $P_\mathrm{fus}$ around the barrier
is due to the sudden change of $s_\mathrm{inj}$.
Compared with experimental data in below-barrier region,
our calculated results are smaller than $P_\mathrm{sym}$ obviously.
It is found that our calculations
with $s_\mathrm{inj}^\mathrm{I}$
overestimate the experimental data while
the results with $s_\mathrm{inj}^\mathrm{II}$ well reproduce the
measurements in above-barrier region.
The differences of $P_\mathrm{fus}$ between the calculations
with $s_\mathrm{inj}^\mathrm{I}$ and
$s_\mathrm{inj}^\mathrm{II}$ become larger from $^{48}$Ca to $^{54}$Cr,
because the height of the inner barrier
is more sensitive to the injection
parameter with the increase of the charge number of CN
\cite{Swiatecki2005_PRC71-014602}.
Our calculations demonstrate that
a precise determination of $s_\mathrm{inj}$
is necessary for achieving a reasonable description
of fusion probability and our method can well reproduce the data
of above-barrier collisions.

By using the experimental data of
fissionlike cross section $\sigma_\mathrm{fis}$ and the upper
limit of fusion probability
$P_\mathrm{sym}$ given in
Ref. \cite{Banerjee2019_PRL122-232503},
the experimental fusion cross section
can be estimated as
$\sigma_\mathrm{fis}P_\mathrm{sym}/S$,
where $S$ is the suppression factor used in Ref. \cite{Banerjee2019_PRL122-232503}.
We compare our results of fusion cross section
$\sigma_\mathrm{fus}$ (cf. Eq. \ref{eq:fus})
with those deduced from measurements
\cite{Banerjee2019_PRL122-232503}
in the bottom panel of Fig. \ref{sig},
although they are not fully equivalent.
For the region $E_\mathrm{c.m.} < V^\mathrm{TDHF}_B+2$ MeV,
the calculations with both $s_\mathrm{inj}^\mathrm{I}$
and $s_\mathrm{inj}^\mathrm{II}$ are smaller than the data
about an order of magnitude
while
for other region,
a good agreement
with is achieved by the calculations with $s_\mathrm{inj}^\mathrm{II}$.
Generally speaking,
$P_\mathrm{fus}$ and $\sigma_\mathrm{fus}$
of below-barrier collisions are
not well reproduced but for
above-barrier ones, our calculations with $s_\mathrm{inj}^\mathrm{II}$
are well consistent with data.
Among these three reaction systems,
the differences between experimental data and
our calculations with $s_\mathrm{inj}^\mathrm{II}$
are largest for $^{54}$Cr+$^{208}$Pb.
This might be
due to that static deformation effects
and dynamic pairing are not included in the present investigation.
In addition, during the TDHF evolution,
the dynamic changes of the shapes for the colliding nuclei influence
the radii $R_\mathrm{T}$ and $R_\mathrm{P}$ at the injection points,
leading to changes of injection parameters that affect fusion probabilities further.
In this work, we use the fitted WS potentials that assumes frozen shapes,
its combination with the threshold energy deduced from TDHF might also be a source of discrepancy.

\section{Summary and perspective}
\label{Sec4}
We present a microscopic
calculations of the fusion probability and the CN formation cross section
by combining the TDHF
with the
coupled-channels approach and the FbD model.
The capture cross section is obtained by performing coupled-channels
calculations, in which the involved parameters of the WS potential are fixed by using
the ground-states properties from the static HF calculations
and capture thresholds determined from TDHF simulations.
The fusion probability is
given by using the FbD model
with the only one input parameter, the injection-point distance,
which is estimated by the time evolution of two colliding nuclei
with the TDHF approach.
We apply our model to the cold-fusion reactions
$^{48}$Ca+$^{208}$Pb,
$^{50}$Ti+$^{208}$Pb,
and $^{54}$Cr+$^{208}$Pb.
The dynamic evolution of the central collisions in both below-
and above-barrier regions are obtained with the effective interaction SLy5.
It is found that
the capture thresholds from TDHF calculations
are in line with those extracted from measurements and
the capture cross sections
can well reproduce the experimental data.
By estimating the injection points with the TDHF approach,
the fusion probabilities and resulted
CN formation cross sections agree with experiments reasonably.

In the present work,
the ground states of the target and projectile nuclei
are all spherical or restricted to be spherical.
For hot-fusion reactions,
it is necessary to take static deformation into account
since most actinide nuclei are deformed in their ground states
and the orientation in the entrance channel also affects
the TDHF dynamic evolution,
the capture cross section,
and the injection point.
The orientation effects can be taken into account in both capture
and fusion processes by rotating the initial wave function
of reactants in TDHF simulation.
This generalization is our next step.
Additionally, it should be mentioned that
the dynamic process of colliding nuclei
is very complicated and the formation of CN from
contact configuration undergoes
complex evolution of shape degrees of freedom.
This challenges the definition of the surfaces
for two colliding nuclei.
Finally, we hope that our microscopic approach
can provide new and reliable supports
for choosing the optimal combination of target and projectile nuclei for
the synthesis of the SHEs with $Z=119$ and $Z=120$ in the future.

\acknowledgments
We thank Liang Li, Yun Huang, Zhen-Ji Wu, Bing Wang,
Bing-Nan Lu, and
Shan-Gui Zhou for helpful discussions.
This work has been supported by
the Strategic Priority Research Program of Chinese Academy of Sciences
(Grant No. XDB34010000 and No. XDPB15) and
the National Natural Science Foundation of China (Grants No. 11975237,
No. 11575189, and No. 11790325).
The results described in this paper are obtained on
the High-performance Computing Cluster of ITP-CAS and
the ScGrid of the Supercomputing Center,
Computer Network Information Center of Chinese Academy of Sciences.


\begin{thebibliography}{91}%
\makeatletter
\providecommand \@ifxundefined [1]{%
 \@ifx{#1\undefined}
}%
\providecommand \@ifnum [1]{%
 \ifnum #1\expandafter \@firstoftwo
 \else \expandafter \@secondoftwo
 \fi
}%
\providecommand \@ifx [1]{%
 \ifx #1\expandafter \@firstoftwo
 \else \expandafter \@secondoftwo
 \fi
}%
\providecommand \natexlab [1]{#1}%
\providecommand \enquote  [1]{``#1''}%
\providecommand \bibnamefont  [1]{#1}%
\providecommand \bibfnamefont [1]{#1}%
\providecommand \citenamefont [1]{#1}%
\providecommand \href@noop [0]{\@secondoftwo}%
\providecommand \href [0]{\begingroup \@sanitize@url \@href}%
\providecommand \@href[1]{\@@startlink{#1}\@@href}%
\providecommand \@@href[1]{\endgroup#1\@@endlink}%
\providecommand \@sanitize@url [0]{\catcode `\\12\catcode `\$12\catcode
  `\&12\catcode `\#12\catcode `\^12\catcode `\_12\catcode `\%12\relax}%
\providecommand \@@startlink[1]{}%
\providecommand \@@endlink[0]{}%
\providecommand \url  [0]{\begingroup\@sanitize@url \@url }%
\providecommand \@url [1]{\endgroup\@href {#1}{\urlprefix }}%
\providecommand \urlprefix  [0]{URL }%
\providecommand \Eprint [0]{\href }%
\providecommand \doibase [0]{https://doi.org/}%
\providecommand \selectlanguage [0]{\@gobble}%
\providecommand \bibinfo  [0]{\@secondoftwo}%
\providecommand \bibfield  [0]{\@secondoftwo}%
\providecommand \translation [1]{[#1]}%
\providecommand \BibitemOpen [0]{}%
\providecommand \bibitemStop [0]{}%
\providecommand \bibitemNoStop [0]{.\EOS\space}%
\providecommand \EOS [0]{\spacefactor3000\relax}%
\providecommand \BibitemShut  [1]{\csname bibitem#1\endcsname}%
\let\auto@bib@innerbib\@empty
\bibitem [{\citenamefont {Hofmann}\ and\ \citenamefont
  {M\"unzenberg}(2000)}]{Hofmann2000_RMP72-733}%
  \BibitemOpen
  \bibfield  {author} {\bibinfo {author} {\bibfnamefont {S.}~\bibnamefont
  {Hofmann}}\ and\ \bibinfo {author} {\bibfnamefont {G.}~\bibnamefont
  {M\"unzenberg}},\ }\bibfield  {title} {\bibinfo {title} {The discovery of the
  heaviest elements},\ }\href {https://doi.org/10.1103/RevModPhys.72.733}
  {\bibfield  {journal} {\bibinfo  {journal} {Rev. Mod. Phys.}\ }\textbf
  {\bibinfo {volume} {72}},\ \bibinfo {pages} {733} (\bibinfo {year}
  {2000})}\BibitemShut {NoStop}%
\bibitem [{\citenamefont {Hamilton}\ \emph {et~al.}(2013)\citenamefont
  {Hamilton}, \citenamefont {Hofmann},\ and\ \citenamefont
  {Oganessian}}]{Hamilton2013_ARNPS63-383}%
  \BibitemOpen
  \bibfield  {author} {\bibinfo {author} {\bibfnamefont {J.~H.}\ \bibnamefont
  {Hamilton}}, \bibinfo {author} {\bibfnamefont {S.}~\bibnamefont {Hofmann}},\
  and\ \bibinfo {author} {\bibfnamefont {Y.~T.}\ \bibnamefont {Oganessian}},\
  }\bibfield  {title} {\bibinfo {title} {Search for superheavy nuclei},\ }\href
  {https://doi.org/10.1146/annurev-nucl-102912-144535} {\bibfield  {journal}
  {\bibinfo  {journal} {Annu. Rev. Nucl. Part. Sci.}\ }\textbf {\bibinfo
  {volume} {63}},\ \bibinfo {pages} {383} (\bibinfo {year} {2013})}\BibitemShut
  {NoStop}%
\bibitem [{\citenamefont {Oganessian}\ and\ \citenamefont
  {Utyonkov}(2015)}]{Oganessian2015_RPP78-036301}%
  \BibitemOpen
  \bibfield  {author} {\bibinfo {author} {\bibfnamefont {Y.~T.}\ \bibnamefont
  {Oganessian}}\ and\ \bibinfo {author} {\bibfnamefont {V.~K.}\ \bibnamefont
  {Utyonkov}},\ }\bibfield  {title} {\bibinfo {title} {Super-heavy element
  research},\ }\href {https://doi.org/10.1088/0034-4885/78/3/036301} {\bibfield
   {journal} {\bibinfo  {journal} {Rep. Prog. Phys.}\ }\textbf {\bibinfo
  {volume} {78}},\ \bibinfo {pages} {036301} (\bibinfo {year}
  {2015})}\BibitemShut {NoStop}%
\bibitem [{\citenamefont {Giuliani}\ \emph {et~al.}(2019)\citenamefont
  {Giuliani}, \citenamefont {Matheson}, \citenamefont {Nazarewicz},
  \citenamefont {Olsen}, \citenamefont {Reinhard}, \citenamefont {Sadhukhan},
  \citenamefont {Schuetrumpf}, \citenamefont {Schunck},\ and\ \citenamefont
  {Schwerdtfeger}}]{Giuliani2019_RMP91-011001}%
  \BibitemOpen
  \bibfield  {author} {\bibinfo {author} {\bibfnamefont {S.~A.}\ \bibnamefont
  {Giuliani}}, \bibinfo {author} {\bibfnamefont {Z.}~\bibnamefont {Matheson}},
  \bibinfo {author} {\bibfnamefont {W.}~\bibnamefont {Nazarewicz}}, \bibinfo
  {author} {\bibfnamefont {E.}~\bibnamefont {Olsen}}, \bibinfo {author}
  {\bibfnamefont {P.-G.}\ \bibnamefont {Reinhard}}, \bibinfo {author}
  {\bibfnamefont {J.}~\bibnamefont {Sadhukhan}}, \bibinfo {author}
  {\bibfnamefont {B.}~\bibnamefont {Schuetrumpf}}, \bibinfo {author}
  {\bibfnamefont {N.}~\bibnamefont {Schunck}},\ and\ \bibinfo {author}
  {\bibfnamefont {P.}~\bibnamefont {Schwerdtfeger}},\ }\bibfield  {title}
  {\bibinfo {title} {Colloquium: Superheavy elements: Oganesson and beyond},\
  }\href {https://doi.org/10.1103/RevModPhys.91.011001} {\bibfield  {journal}
  {\bibinfo  {journal} {Rev. Mod. Phys.}\ }\textbf {\bibinfo {volume} {91}},\
  \bibinfo {pages} {011001} (\bibinfo {year} {2019})}\BibitemShut {NoStop}%
\bibitem [{\citenamefont {Morita}\ \emph {et~al.}(2004)\citenamefont {Morita},
  \citenamefont {Morimoto}, \citenamefont {Kaji}, \citenamefont {Akiyama},
  \citenamefont {Goto}, \citenamefont {Haba}, \citenamefont {Ideguchi},
  \citenamefont {Kanungo}, \citenamefont {Katori}, \citenamefont {Koura},
  \citenamefont {Kudo}, \citenamefont {Ohnishi}, \citenamefont {Ozawa},
  \citenamefont {Suda}, \citenamefont {Sueki}, \citenamefont {Xu},
  \citenamefont {Yamaguchi}, \citenamefont {Yoneda}, \citenamefont {Yoshida},\
  and\ \citenamefont {Zhao}}]{Morita2004_JPSJ73-2593}%
  \BibitemOpen
  \bibfield  {author} {\bibinfo {author} {\bibfnamefont {K.}~\bibnamefont
  {Morita}}, \bibinfo {author} {\bibfnamefont {K.}~\bibnamefont {Morimoto}},
  \bibinfo {author} {\bibfnamefont {D.}~\bibnamefont {Kaji}}, \bibinfo {author}
  {\bibfnamefont {T.}~\bibnamefont {Akiyama}}, \bibinfo {author} {\bibfnamefont
  {S.-i.}\ \bibnamefont {Goto}}, \bibinfo {author} {\bibfnamefont
  {H.}~\bibnamefont {Haba}}, \bibinfo {author} {\bibfnamefont {E.}~\bibnamefont
  {Ideguchi}}, \bibinfo {author} {\bibfnamefont {R.}~\bibnamefont {Kanungo}},
  \bibinfo {author} {\bibfnamefont {K.}~\bibnamefont {Katori}}, \bibinfo
  {author} {\bibfnamefont {H.}~\bibnamefont {Koura}}, \bibinfo {author}
  {\bibfnamefont {H.}~\bibnamefont {Kudo}}, \bibinfo {author} {\bibfnamefont
  {T.}~\bibnamefont {Ohnishi}}, \bibinfo {author} {\bibfnamefont
  {A.}~\bibnamefont {Ozawa}}, \bibinfo {author} {\bibfnamefont
  {T.}~\bibnamefont {Suda}}, \bibinfo {author} {\bibfnamefont {K.}~\bibnamefont
  {Sueki}}, \bibinfo {author} {\bibfnamefont {H.-S.}\ \bibnamefont {Xu}},
  \bibinfo {author} {\bibfnamefont {T.}~\bibnamefont {Yamaguchi}}, \bibinfo
  {author} {\bibfnamefont {A.}~\bibnamefont {Yoneda}}, \bibinfo {author}
  {\bibfnamefont {A.}~\bibnamefont {Yoshida}},\ and\ \bibinfo {author}
  {\bibfnamefont {Y.-L.}\ \bibnamefont {Zhao}},\ }\bibfield  {title} {\bibinfo
  {title} {Experiment on the synthesis of element 113 in the reaction
  $^{209}${B}i($^{70}${Z}n,n)$^{278}$113},\ }\href
  {https://doi.org/10.1143/JPSJ.73.2593} {\bibfield  {journal} {\bibinfo
  {journal} {J. Phys. Soc. Jpn.}\ }\textbf {\bibinfo {volume} {73}},\ \bibinfo
  {pages} {2593} (\bibinfo {year} {2004})}\BibitemShut {NoStop}%
\bibitem [{\citenamefont {Oganessian}(2007)}]{Oganessian2007_JPG34-R165}%
  \BibitemOpen
  \bibfield  {author} {\bibinfo {author} {\bibfnamefont {Y.}~\bibnamefont
  {Oganessian}},\ }\bibfield  {title} {\bibinfo {title} {{Heaviest nuclei from
  $^{48}${C}a-induced reactions}},\ }\href
  {https://doi.org/10.1088/0954-3899/34/4/r01} {\bibfield  {journal} {\bibinfo
  {journal} {J. Phys. G: Nucl. Part. Phys.}\ }\textbf {\bibinfo {volume}
  {34}},\ \bibinfo {pages} {R165} (\bibinfo {year} {2007})}\BibitemShut
  {NoStop}%
\bibitem [{\citenamefont {Oganessian}\ \emph {et~al.}(2010)\citenamefont
  {Oganessian}, \citenamefont {Abdullin}, \citenamefont {Bailey}, \citenamefont
  {Benker}, \citenamefont {Bennett}, \citenamefont {Dmitriev}, \citenamefont
  {Ezold}, \citenamefont {Hamilton}, \citenamefont {Henderson}, \citenamefont
  {Itkis}, \citenamefont {Lobanov}, \citenamefont {Mezentsev}, \citenamefont
  {Moody}, \citenamefont {Nelson}, \citenamefont {Polyakov}, \citenamefont
  {Porter}, \citenamefont {Ramayya}, \citenamefont {Riley}, \citenamefont
  {Roberto}, \citenamefont {Ryabinin}, \citenamefont {Rykaczewski},
  \citenamefont {Sagaidak}, \citenamefont {Shaughnessy}, \citenamefont
  {Shirokovsky}, \citenamefont {Stoyer}, \citenamefont {Subbotin},
  \citenamefont {Sudowe}, \citenamefont {Sukhov}, \citenamefont {Tsyganov},
  \citenamefont {Utyonkov}, \citenamefont {Voinov}, \citenamefont {Vostokin},\
  and\ \citenamefont {Wilk}}]{Oganessian2010_PRL104-142502}%
  \BibitemOpen
  \bibfield  {author} {\bibinfo {author} {\bibfnamefont {Y.~T.}\ \bibnamefont
  {Oganessian}}, \bibinfo {author} {\bibfnamefont {F.~S.}\ \bibnamefont
  {Abdullin}}, \bibinfo {author} {\bibfnamefont {P.~D.}\ \bibnamefont
  {Bailey}}, \bibinfo {author} {\bibfnamefont {D.~E.}\ \bibnamefont {Benker}},
  \bibinfo {author} {\bibfnamefont {M.~E.}\ \bibnamefont {Bennett}}, \bibinfo
  {author} {\bibfnamefont {S.~N.}\ \bibnamefont {Dmitriev}}, \bibinfo {author}
  {\bibfnamefont {J.~G.}\ \bibnamefont {Ezold}}, \bibinfo {author}
  {\bibfnamefont {J.~H.}\ \bibnamefont {Hamilton}}, \bibinfo {author}
  {\bibfnamefont {R.~A.}\ \bibnamefont {Henderson}}, \bibinfo {author}
  {\bibfnamefont {M.~G.}\ \bibnamefont {Itkis}}, \bibinfo {author}
  {\bibfnamefont {Y.~V.}\ \bibnamefont {Lobanov}}, \bibinfo {author}
  {\bibfnamefont {A.~N.}\ \bibnamefont {Mezentsev}}, \bibinfo {author}
  {\bibfnamefont {K.~J.}\ \bibnamefont {Moody}}, \bibinfo {author}
  {\bibfnamefont {S.~L.}\ \bibnamefont {Nelson}}, \bibinfo {author}
  {\bibfnamefont {A.~N.}\ \bibnamefont {Polyakov}}, \bibinfo {author}
  {\bibfnamefont {C.~E.}\ \bibnamefont {Porter}}, \bibinfo {author}
  {\bibfnamefont {A.~V.}\ \bibnamefont {Ramayya}}, \bibinfo {author}
  {\bibfnamefont {F.~D.}\ \bibnamefont {Riley}}, \bibinfo {author}
  {\bibfnamefont {J.~B.}\ \bibnamefont {Roberto}}, \bibinfo {author}
  {\bibfnamefont {M.~A.}\ \bibnamefont {Ryabinin}}, \bibinfo {author}
  {\bibfnamefont {K.~P.}\ \bibnamefont {Rykaczewski}}, \bibinfo {author}
  {\bibfnamefont {R.~N.}\ \bibnamefont {Sagaidak}}, \bibinfo {author}
  {\bibfnamefont {D.~A.}\ \bibnamefont {Shaughnessy}}, \bibinfo {author}
  {\bibfnamefont {I.~V.}\ \bibnamefont {Shirokovsky}}, \bibinfo {author}
  {\bibfnamefont {M.~A.}\ \bibnamefont {Stoyer}}, \bibinfo {author}
  {\bibfnamefont {V.~G.}\ \bibnamefont {Subbotin}}, \bibinfo {author}
  {\bibfnamefont {R.}~\bibnamefont {Sudowe}}, \bibinfo {author} {\bibfnamefont
  {A.~M.}\ \bibnamefont {Sukhov}}, \bibinfo {author} {\bibfnamefont {Y.~S.}\
  \bibnamefont {Tsyganov}}, \bibinfo {author} {\bibfnamefont {V.~K.}\
  \bibnamefont {Utyonkov}}, \bibinfo {author} {\bibfnamefont {A.~A.}\
  \bibnamefont {Voinov}}, \bibinfo {author} {\bibfnamefont {G.~K.}\
  \bibnamefont {Vostokin}},\ and\ \bibinfo {author} {\bibfnamefont {P.~A.}\
  \bibnamefont {Wilk}},\ }\bibfield  {title} {\bibinfo {title} {Synthesis of a
  new element with atomic number ${Z}=117$},\ }\href
  {https://doi.org/10.1103/PhysRevLett.104.142502} {\bibfield  {journal}
  {\bibinfo  {journal} {Phys. Rev. Lett.}\ }\textbf {\bibinfo {volume} {104}},\
  \bibinfo {pages} {142502} (\bibinfo {year} {2010})}\BibitemShut {NoStop}%
\bibitem [{\citenamefont {Oganessian}\ \emph {et~al.}(2009)\citenamefont
  {Oganessian}, \citenamefont {Utyonkov}, \citenamefont {Lobanov},
  \citenamefont {Abdullin}, \citenamefont {Polyakov}, \citenamefont {Sagaidak},
  \citenamefont {Shirokovsky}, \citenamefont {Tsyganov}, \citenamefont
  {Voinov}, \citenamefont {Mezentsev}, \citenamefont {Subbotin}, \citenamefont
  {Sukhov}, \citenamefont {Subotic}, \citenamefont {Zagrebaev}, \citenamefont
  {Dmitriev}, \citenamefont {Henderson}, \citenamefont {Moody}, \citenamefont
  {Kenneally}, \citenamefont {Landrum}, \citenamefont {Shaughnessy},
  \citenamefont {Stoyer}, \citenamefont {Stoyer},\ and\ \citenamefont
  {Wilk}}]{Oganessian2009_PRC79-024603}%
  \BibitemOpen
  \bibfield  {author} {\bibinfo {author} {\bibfnamefont {Y.~T.}\ \bibnamefont
  {Oganessian}}, \bibinfo {author} {\bibfnamefont {V.~K.}\ \bibnamefont
  {Utyonkov}}, \bibinfo {author} {\bibfnamefont {Y.~V.}\ \bibnamefont
  {Lobanov}}, \bibinfo {author} {\bibfnamefont {F.~S.}\ \bibnamefont
  {Abdullin}}, \bibinfo {author} {\bibfnamefont {A.~N.}\ \bibnamefont
  {Polyakov}}, \bibinfo {author} {\bibfnamefont {R.~N.}\ \bibnamefont
  {Sagaidak}}, \bibinfo {author} {\bibfnamefont {I.~V.}\ \bibnamefont
  {Shirokovsky}}, \bibinfo {author} {\bibfnamefont {Y.~S.}\ \bibnamefont
  {Tsyganov}}, \bibinfo {author} {\bibfnamefont {A.~A.}\ \bibnamefont
  {Voinov}}, \bibinfo {author} {\bibfnamefont {A.~N.}\ \bibnamefont
  {Mezentsev}}, \bibinfo {author} {\bibfnamefont {V.~G.}\ \bibnamefont
  {Subbotin}}, \bibinfo {author} {\bibfnamefont {A.~M.}\ \bibnamefont
  {Sukhov}}, \bibinfo {author} {\bibfnamefont {K.}~\bibnamefont {Subotic}},
  \bibinfo {author} {\bibfnamefont {V.~I.}\ \bibnamefont {Zagrebaev}}, \bibinfo
  {author} {\bibfnamefont {S.~N.}\ \bibnamefont {Dmitriev}}, \bibinfo {author}
  {\bibfnamefont {R.~A.}\ \bibnamefont {Henderson}}, \bibinfo {author}
  {\bibfnamefont {K.~J.}\ \bibnamefont {Moody}}, \bibinfo {author}
  {\bibfnamefont {J.~M.}\ \bibnamefont {Kenneally}}, \bibinfo {author}
  {\bibfnamefont {J.~H.}\ \bibnamefont {Landrum}}, \bibinfo {author}
  {\bibfnamefont {D.~A.}\ \bibnamefont {Shaughnessy}}, \bibinfo {author}
  {\bibfnamefont {M.~A.}\ \bibnamefont {Stoyer}}, \bibinfo {author}
  {\bibfnamefont {N.~J.}\ \bibnamefont {Stoyer}},\ and\ \bibinfo {author}
  {\bibfnamefont {P.~A.}\ \bibnamefont {Wilk}},\ }\bibfield  {title} {\bibinfo
  {title} {Attempt to produce element 120 in the
  $^{244}\mathrm{Pu}+{}^{58}\mathrm{Fe}$ reaction},\ }\href
  {https://doi.org/10.1103/PhysRevC.79.024603} {\bibfield  {journal} {\bibinfo
  {journal} {Phys. Rev. C}\ }\textbf {\bibinfo {volume} {79}},\ \bibinfo
  {pages} {024603} (\bibinfo {year} {2009})}\BibitemShut {NoStop}%
\bibitem [{\citenamefont {Kozulin}\ \emph {et~al.}(2010)\citenamefont
  {Kozulin}, \citenamefont {Knyazheva}, \citenamefont {Itkis}, \citenamefont
  {Itkis}, \citenamefont {Bogachev}, \citenamefont {Krupa}, \citenamefont
  {Loktev}, \citenamefont {Smirnov}, \citenamefont {Zagrebaev}, \citenamefont
  {{\ifmmode\ddot{A}\else\"{A}\fi}yst{\ifmmode\ddot{o}\else\"{o}\fi}},
  \citenamefont {Trzaska}, \citenamefont {Rubchenya}, \citenamefont {Vardaci},
  \citenamefont {Stefanini}, \citenamefont {Cinausero}, \citenamefont
  {Corradi}, \citenamefont {Fioretto}, \citenamefont {Mason}, \citenamefont
  {Prete}, \citenamefont {Silvestri}, \citenamefont {Beghini}, \citenamefont
  {Montagnoli}, \citenamefont {Scarlassara}, \citenamefont {Hanappe},
  \citenamefont {Khlebnikov}, \citenamefont {Kliman}, \citenamefont {Brondi},
  \citenamefont {Di~Nitto}, \citenamefont {Moro}, \citenamefont {Gelli},\ and\
  \citenamefont {Szilner}}]{Kozulin2010_PLB686-227}%
  \BibitemOpen
  \bibfield  {author} {\bibinfo {author} {\bibfnamefont {E.~M.}\ \bibnamefont
  {Kozulin}}, \bibinfo {author} {\bibfnamefont {G.~N.}\ \bibnamefont
  {Knyazheva}}, \bibinfo {author} {\bibfnamefont {I.~M.}\ \bibnamefont
  {Itkis}}, \bibinfo {author} {\bibfnamefont {M.~G.}\ \bibnamefont {Itkis}},
  \bibinfo {author} {\bibfnamefont {A.~A.}\ \bibnamefont {Bogachev}}, \bibinfo
  {author} {\bibfnamefont {L.}~\bibnamefont {Krupa}}, \bibinfo {author}
  {\bibfnamefont {T.~A.}\ \bibnamefont {Loktev}}, \bibinfo {author}
  {\bibfnamefont {S.~V.}\ \bibnamefont {Smirnov}}, \bibinfo {author}
  {\bibfnamefont {V.~I.}\ \bibnamefont {Zagrebaev}}, \bibinfo {author}
  {\bibfnamefont {J.}~\bibnamefont
  {{\ifmmode\ddot{A}\else\"{A}\fi}yst{\ifmmode\ddot{o}\else\"{o}\fi}}},
  \bibinfo {author} {\bibfnamefont {W.~H.}\ \bibnamefont {Trzaska}}, \bibinfo
  {author} {\bibfnamefont {V.~A.}\ \bibnamefont {Rubchenya}}, \bibinfo {author}
  {\bibfnamefont {E.}~\bibnamefont {Vardaci}}, \bibinfo {author} {\bibfnamefont
  {A.~M.}\ \bibnamefont {Stefanini}}, \bibinfo {author} {\bibfnamefont
  {M.}~\bibnamefont {Cinausero}}, \bibinfo {author} {\bibfnamefont
  {L.}~\bibnamefont {Corradi}}, \bibinfo {author} {\bibfnamefont
  {E.}~\bibnamefont {Fioretto}}, \bibinfo {author} {\bibfnamefont
  {P.}~\bibnamefont {Mason}}, \bibinfo {author} {\bibfnamefont {G.~F.}\
  \bibnamefont {Prete}}, \bibinfo {author} {\bibfnamefont {R.}~\bibnamefont
  {Silvestri}}, \bibinfo {author} {\bibfnamefont {S.}~\bibnamefont {Beghini}},
  \bibinfo {author} {\bibfnamefont {G.}~\bibnamefont {Montagnoli}}, \bibinfo
  {author} {\bibfnamefont {F.}~\bibnamefont {Scarlassara}}, \bibinfo {author}
  {\bibfnamefont {F.}~\bibnamefont {Hanappe}}, \bibinfo {author} {\bibfnamefont
  {S.~V.}\ \bibnamefont {Khlebnikov}}, \bibinfo {author} {\bibfnamefont
  {J.}~\bibnamefont {Kliman}}, \bibinfo {author} {\bibfnamefont
  {A.}~\bibnamefont {Brondi}}, \bibinfo {author} {\bibfnamefont
  {A.}~\bibnamefont {Di~Nitto}}, \bibinfo {author} {\bibfnamefont
  {R.}~\bibnamefont {Moro}}, \bibinfo {author} {\bibfnamefont {N.}~\bibnamefont
  {Gelli}},\ and\ \bibinfo {author} {\bibfnamefont {S.}~\bibnamefont
  {Szilner}},\ }\bibfield  {title} {\bibinfo {title} {Investigation of the
  reaction $^{64}${N}i+$^{238}${U} being an option of synthesizing element
  120},\ }\href {https://doi.org/10.1016/j.physletb.2010.02.041} {\bibfield
  {journal} {\bibinfo  {journal} {Phys. Lett. B}\ }\textbf {\bibinfo {volume}
  {686}},\ \bibinfo {pages} {227} (\bibinfo {year} {2010})}\BibitemShut
  {NoStop}%
\bibitem [{\citenamefont {Hofmann}\ \emph {et~al.}(2016)\citenamefont
  {Hofmann}, \citenamefont {Heinz}, \citenamefont {Mann}, \citenamefont
  {Maurer}, \citenamefont {M{\ifmmode\ddot{u}\else\"{u}\fi}nzenberg},
  \citenamefont {Antalic}, \citenamefont {Barth}, \citenamefont {Burkhard},
  \citenamefont {Dahl}, \citenamefont {Eberhardt}, \citenamefont {Grzywacz},
  \citenamefont {Hamilton}, \citenamefont {Henderson}, \citenamefont
  {Kenneally}, \citenamefont {Kindler}, \citenamefont {Kojouharov},
  \citenamefont {Lang}, \citenamefont {Lommel}, \citenamefont {Miernik},
  \citenamefont {Miller}, \citenamefont {Moody}, \citenamefont {Morita},
  \citenamefont {Nishio}, \citenamefont {Popeko}, \citenamefont {Roberto},
  \citenamefont {Runke}, \citenamefont {Rykaczewski}, \citenamefont {Saro},
  \citenamefont {Scheidenberger}, \citenamefont
  {Sch{\ifmmode\ddot{o}\else\"{o}\fi}tt}, \citenamefont {Shaughnessy},
  \citenamefont {Stoyer}, \citenamefont
  {Th{\ifmmode\ddot{o}\else\"{o}\fi}rle-Pospiech}, \citenamefont {Tinschert},
  \citenamefont {Trautmann}, \citenamefont {Uusitalo},\ and\ \citenamefont
  {Yeremin}}]{Hofmann2016_EPJA52-180}%
  \BibitemOpen
  \bibfield  {author} {\bibinfo {author} {\bibfnamefont {S.}~\bibnamefont
  {Hofmann}}, \bibinfo {author} {\bibfnamefont {S.}~\bibnamefont {Heinz}},
  \bibinfo {author} {\bibfnamefont {R.}~\bibnamefont {Mann}}, \bibinfo {author}
  {\bibfnamefont {J.}~\bibnamefont {Maurer}}, \bibinfo {author} {\bibfnamefont
  {G.}~\bibnamefont {M{\ifmmode\ddot{u}\else\"{u}\fi}nzenberg}}, \bibinfo
  {author} {\bibfnamefont {S.}~\bibnamefont {Antalic}}, \bibinfo {author}
  {\bibfnamefont {W.}~\bibnamefont {Barth}}, \bibinfo {author} {\bibfnamefont
  {H.~G.}\ \bibnamefont {Burkhard}}, \bibinfo {author} {\bibfnamefont
  {L.}~\bibnamefont {Dahl}}, \bibinfo {author} {\bibfnamefont {K.}~\bibnamefont
  {Eberhardt}}, \bibinfo {author} {\bibfnamefont {R.}~\bibnamefont {Grzywacz}},
  \bibinfo {author} {\bibfnamefont {J.~H.}\ \bibnamefont {Hamilton}}, \bibinfo
  {author} {\bibfnamefont {R.~A.}\ \bibnamefont {Henderson}}, \bibinfo {author}
  {\bibfnamefont {J.~M.}\ \bibnamefont {Kenneally}}, \bibinfo {author}
  {\bibfnamefont {B.}~\bibnamefont {Kindler}}, \bibinfo {author} {\bibfnamefont
  {I.}~\bibnamefont {Kojouharov}}, \bibinfo {author} {\bibfnamefont
  {R.}~\bibnamefont {Lang}}, \bibinfo {author} {\bibfnamefont {B.}~\bibnamefont
  {Lommel}}, \bibinfo {author} {\bibfnamefont {K.}~\bibnamefont {Miernik}},
  \bibinfo {author} {\bibfnamefont {D.}~\bibnamefont {Miller}}, \bibinfo
  {author} {\bibfnamefont {K.~J.}\ \bibnamefont {Moody}}, \bibinfo {author}
  {\bibfnamefont {K.}~\bibnamefont {Morita}}, \bibinfo {author} {\bibfnamefont
  {K.}~\bibnamefont {Nishio}}, \bibinfo {author} {\bibfnamefont {A.~G.}\
  \bibnamefont {Popeko}}, \bibinfo {author} {\bibfnamefont {J.~B.}\
  \bibnamefont {Roberto}}, \bibinfo {author} {\bibfnamefont {J.}~\bibnamefont
  {Runke}}, \bibinfo {author} {\bibfnamefont {K.~P.}\ \bibnamefont
  {Rykaczewski}}, \bibinfo {author} {\bibfnamefont {S.}~\bibnamefont {Saro}},
  \bibinfo {author} {\bibfnamefont {C.}~\bibnamefont {Scheidenberger}},
  \bibinfo {author} {\bibfnamefont {H.~J.}\ \bibnamefont
  {Sch{\ifmmode\ddot{o}\else\"{o}\fi}tt}}, \bibinfo {author} {\bibfnamefont
  {D.~A.}\ \bibnamefont {Shaughnessy}}, \bibinfo {author} {\bibfnamefont
  {M.~A.}\ \bibnamefont {Stoyer}}, \bibinfo {author} {\bibfnamefont
  {P.}~\bibnamefont {Th{\ifmmode\ddot{o}\else\"{o}\fi}rle-Pospiech}}, \bibinfo
  {author} {\bibfnamefont {K.}~\bibnamefont {Tinschert}}, \bibinfo {author}
  {\bibfnamefont {N.}~\bibnamefont {Trautmann}}, \bibinfo {author}
  {\bibfnamefont {J.}~\bibnamefont {Uusitalo}},\ and\ \bibinfo {author}
  {\bibfnamefont {A.~V.}\ \bibnamefont {Yeremin}},\ }\bibfield  {title}
  {\bibinfo {title} {{Review of even element super-heavy nuclei and search for
  element 120}},\ }\href {https://doi.org/10.1140/epja/i2016-16180-4}
  {\bibfield  {journal} {\bibinfo  {journal} {Eur. Phys. J. A}\ }\textbf
  {\bibinfo {volume} {52}},\ \bibinfo {pages} {180} (\bibinfo {year}
  {2016})}\BibitemShut {NoStop}%
\bibitem [{\citenamefont {Khuyagbaatar}\ \emph {et~al.}(2020)\citenamefont
  {Khuyagbaatar}, \citenamefont {Yakushev}, \citenamefont {D\"ullmann},
  \citenamefont {Ackermann}, \citenamefont {Andersson}, \citenamefont {Asai},
  \citenamefont {Block}, \citenamefont {Boll}, \citenamefont {Brand},
  \citenamefont {Cox}, \citenamefont {Dasgupta}, \citenamefont {Derkx},
  \citenamefont {Di~Nitto}, \citenamefont {Eberhardt}, \citenamefont {Even},
  \citenamefont {Evers}, \citenamefont {Fahlander}, \citenamefont {Forsberg},
  \citenamefont {Gates}, \citenamefont {Gharibyan}, \citenamefont {Golubev},
  \citenamefont {Gregorich}, \citenamefont {Hamilton}, \citenamefont
  {Hartmann}, \citenamefont {Herzberg}, \citenamefont {He\ss{}berger},
  \citenamefont {Hinde}, \citenamefont {Hoffmann}, \citenamefont {Hollinger},
  \citenamefont {H\"ubner}, \citenamefont {J\"ager}, \citenamefont {Kindler},
  \citenamefont {Kratz}, \citenamefont {Krier}, \citenamefont {Kurz},
  \citenamefont {Laatiaoui}, \citenamefont {Lahiri}, \citenamefont {Lang},
  \citenamefont {Lommel}, \citenamefont {Maiti}, \citenamefont {Miernik},
  \citenamefont {Minami}, \citenamefont {Mistry}, \citenamefont {Mokry},
  \citenamefont {Nitsche}, \citenamefont {Omtvedt}, \citenamefont {Pang},
  \citenamefont {Papadakis}, \citenamefont {Renisch}, \citenamefont {Roberto},
  \citenamefont {Rudolph}, \citenamefont {Runke}, \citenamefont {Rykaczewski},
  \citenamefont {Sarmiento}, \citenamefont {Sch\"adel}, \citenamefont
  {Schausten}, \citenamefont {Semchenkov}, \citenamefont {Shaughnessy},
  \citenamefont {Steinegger}, \citenamefont {Steiner}, \citenamefont
  {Tereshatov}, \citenamefont {Th\"orle-Pospiech}, \citenamefont {Tinschert},
  \citenamefont {Torres De~Heidenreich}, \citenamefont {Trautmann},
  \citenamefont {T\"urler}, \citenamefont {Uusitalo}, \citenamefont
  {Wegrzecki}, \citenamefont {Wiehl}, \citenamefont {Van~Cleve},\ and\
  \citenamefont {Yakusheva}}]{Khuyagbaatar2020_PRC102-064602}%
  \BibitemOpen
  \bibfield  {author} {\bibinfo {author} {\bibfnamefont {J.}~\bibnamefont
  {Khuyagbaatar}}, \bibinfo {author} {\bibfnamefont {A.}~\bibnamefont
  {Yakushev}}, \bibinfo {author} {\bibfnamefont {C.~E.}\ \bibnamefont
  {D\"ullmann}}, \bibinfo {author} {\bibfnamefont {D.}~\bibnamefont
  {Ackermann}}, \bibinfo {author} {\bibfnamefont {L.-L.}\ \bibnamefont
  {Andersson}}, \bibinfo {author} {\bibfnamefont {M.}~\bibnamefont {Asai}},
  \bibinfo {author} {\bibfnamefont {M.}~\bibnamefont {Block}}, \bibinfo
  {author} {\bibfnamefont {R.~A.}\ \bibnamefont {Boll}}, \bibinfo {author}
  {\bibfnamefont {H.}~\bibnamefont {Brand}}, \bibinfo {author} {\bibfnamefont
  {D.~M.}\ \bibnamefont {Cox}}, \bibinfo {author} {\bibfnamefont
  {M.}~\bibnamefont {Dasgupta}}, \bibinfo {author} {\bibfnamefont
  {X.}~\bibnamefont {Derkx}}, \bibinfo {author} {\bibfnamefont
  {A.}~\bibnamefont {Di~Nitto}}, \bibinfo {author} {\bibfnamefont
  {K.}~\bibnamefont {Eberhardt}}, \bibinfo {author} {\bibfnamefont
  {J.}~\bibnamefont {Even}}, \bibinfo {author} {\bibfnamefont {M.}~\bibnamefont
  {Evers}}, \bibinfo {author} {\bibfnamefont {C.}~\bibnamefont {Fahlander}},
  \bibinfo {author} {\bibfnamefont {U.}~\bibnamefont {Forsberg}}, \bibinfo
  {author} {\bibfnamefont {J.~M.}\ \bibnamefont {Gates}}, \bibinfo {author}
  {\bibfnamefont {N.}~\bibnamefont {Gharibyan}}, \bibinfo {author}
  {\bibfnamefont {P.}~\bibnamefont {Golubev}}, \bibinfo {author} {\bibfnamefont
  {K.~E.}\ \bibnamefont {Gregorich}}, \bibinfo {author} {\bibfnamefont {J.~H.}\
  \bibnamefont {Hamilton}}, \bibinfo {author} {\bibfnamefont {W.}~\bibnamefont
  {Hartmann}}, \bibinfo {author} {\bibfnamefont {R.-D.}\ \bibnamefont
  {Herzberg}}, \bibinfo {author} {\bibfnamefont {F.~P.}\ \bibnamefont
  {He\ss{}berger}}, \bibinfo {author} {\bibfnamefont {D.~J.}\ \bibnamefont
  {Hinde}}, \bibinfo {author} {\bibfnamefont {J.}~\bibnamefont {Hoffmann}},
  \bibinfo {author} {\bibfnamefont {R.}~\bibnamefont {Hollinger}}, \bibinfo
  {author} {\bibfnamefont {A.}~\bibnamefont {H\"ubner}}, \bibinfo {author}
  {\bibfnamefont {E.}~\bibnamefont {J\"ager}}, \bibinfo {author} {\bibfnamefont
  {B.}~\bibnamefont {Kindler}}, \bibinfo {author} {\bibfnamefont {J.~V.}\
  \bibnamefont {Kratz}}, \bibinfo {author} {\bibfnamefont {J.}~\bibnamefont
  {Krier}}, \bibinfo {author} {\bibfnamefont {N.}~\bibnamefont {Kurz}},
  \bibinfo {author} {\bibfnamefont {M.}~\bibnamefont {Laatiaoui}}, \bibinfo
  {author} {\bibfnamefont {S.}~\bibnamefont {Lahiri}}, \bibinfo {author}
  {\bibfnamefont {R.}~\bibnamefont {Lang}}, \bibinfo {author} {\bibfnamefont
  {B.}~\bibnamefont {Lommel}}, \bibinfo {author} {\bibfnamefont
  {M.}~\bibnamefont {Maiti}}, \bibinfo {author} {\bibfnamefont
  {K.}~\bibnamefont {Miernik}}, \bibinfo {author} {\bibfnamefont
  {S.}~\bibnamefont {Minami}}, \bibinfo {author} {\bibfnamefont {A.~K.}\
  \bibnamefont {Mistry}}, \bibinfo {author} {\bibfnamefont {C.}~\bibnamefont
  {Mokry}}, \bibinfo {author} {\bibfnamefont {H.}~\bibnamefont {Nitsche}},
  \bibinfo {author} {\bibfnamefont {J.~P.}\ \bibnamefont {Omtvedt}}, \bibinfo
  {author} {\bibfnamefont {G.~K.}\ \bibnamefont {Pang}}, \bibinfo {author}
  {\bibfnamefont {P.}~\bibnamefont {Papadakis}}, \bibinfo {author}
  {\bibfnamefont {D.}~\bibnamefont {Renisch}}, \bibinfo {author} {\bibfnamefont
  {J.~B.}\ \bibnamefont {Roberto}}, \bibinfo {author} {\bibfnamefont
  {D.}~\bibnamefont {Rudolph}}, \bibinfo {author} {\bibfnamefont
  {J.}~\bibnamefont {Runke}}, \bibinfo {author} {\bibfnamefont {K.~P.}\
  \bibnamefont {Rykaczewski}}, \bibinfo {author} {\bibfnamefont {L.~G.}\
  \bibnamefont {Sarmiento}}, \bibinfo {author} {\bibfnamefont {M.}~\bibnamefont
  {Sch\"adel}}, \bibinfo {author} {\bibfnamefont {B.}~\bibnamefont
  {Schausten}}, \bibinfo {author} {\bibfnamefont {A.}~\bibnamefont
  {Semchenkov}}, \bibinfo {author} {\bibfnamefont {D.~A.}\ \bibnamefont
  {Shaughnessy}}, \bibinfo {author} {\bibfnamefont {P.}~\bibnamefont
  {Steinegger}}, \bibinfo {author} {\bibfnamefont {J.}~\bibnamefont {Steiner}},
  \bibinfo {author} {\bibfnamefont {E.~E.}\ \bibnamefont {Tereshatov}},
  \bibinfo {author} {\bibfnamefont {P.}~\bibnamefont {Th\"orle-Pospiech}},
  \bibinfo {author} {\bibfnamefont {K.}~\bibnamefont {Tinschert}}, \bibinfo
  {author} {\bibfnamefont {T.}~\bibnamefont {Torres De~Heidenreich}}, \bibinfo
  {author} {\bibfnamefont {N.}~\bibnamefont {Trautmann}}, \bibinfo {author}
  {\bibfnamefont {A.}~\bibnamefont {T\"urler}}, \bibinfo {author}
  {\bibfnamefont {J.}~\bibnamefont {Uusitalo}}, \bibinfo {author}
  {\bibfnamefont {M.}~\bibnamefont {Wegrzecki}}, \bibinfo {author}
  {\bibfnamefont {N.}~\bibnamefont {Wiehl}}, \bibinfo {author} {\bibfnamefont
  {S.~M.}\ \bibnamefont {Van~Cleve}},\ and\ \bibinfo {author} {\bibfnamefont
  {V.}~\bibnamefont {Yakusheva}},\ }\bibfield  {title} {\bibinfo {title}
  {Search for elements 119 and 120},\ }\href
  {https://doi.org/10.1103/PhysRevC.102.064602} {\bibfield  {journal} {\bibinfo
   {journal} {Phys. Rev. C}\ }\textbf {\bibinfo {volume} {102}},\ \bibinfo
  {pages} {064602} (\bibinfo {year} {2020})}\BibitemShut {NoStop}%
\bibitem [{\citenamefont {Albers}\ \emph {et~al.}(2020)\citenamefont {Albers},
  \citenamefont {Khuyagbaatar}, \citenamefont {Hinde}, \citenamefont {Carter},
  \citenamefont {Cook}, \citenamefont {Dasgupta}, \citenamefont {D\"{u}llmann},
  \citenamefont {Eberhardt}, \citenamefont {Jeung}, \citenamefont {Kalkal},
  \citenamefont {Kindler}, \citenamefont {Lobanov}, \citenamefont {Lommel},
  \citenamefont {Mokry}, \citenamefont {Prasad}, \citenamefont {Rafferty},
  \citenamefont {Runke}, \citenamefont {Sekizawa}, \citenamefont {Sengupta},
  \citenamefont {Simenel}, \citenamefont {Simpson}, \citenamefont {Smith},
  \citenamefont {Th\"{o}rle-Pospiech}, \citenamefont {Trautmann}, \citenamefont
  {Vo-Phuoc}, \citenamefont {Walshe}, \citenamefont {Williams},\ and\
  \citenamefont {Yakushev}}]{Albers2020_PLB808-135626}%
  \BibitemOpen
  \bibfield  {author} {\bibinfo {author} {\bibfnamefont {H.~M.}\ \bibnamefont
  {Albers}}, \bibinfo {author} {\bibfnamefont {J.}~\bibnamefont
  {Khuyagbaatar}}, \bibinfo {author} {\bibfnamefont {D.~J.}\ \bibnamefont
  {Hinde}}, \bibinfo {author} {\bibfnamefont {I.~P.}\ \bibnamefont {Carter}},
  \bibinfo {author} {\bibfnamefont {K.~J.}\ \bibnamefont {Cook}}, \bibinfo
  {author} {\bibfnamefont {M.}~\bibnamefont {Dasgupta}}, \bibinfo {author}
  {\bibfnamefont {{\relax Ch}.~E.}\ \bibnamefont {D\"{u}llmann}}, \bibinfo
  {author} {\bibfnamefont {K.}~\bibnamefont {Eberhardt}}, \bibinfo {author}
  {\bibfnamefont {D.~Y.}\ \bibnamefont {Jeung}}, \bibinfo {author}
  {\bibfnamefont {S.}~\bibnamefont {Kalkal}}, \bibinfo {author} {\bibfnamefont
  {B.}~\bibnamefont {Kindler}}, \bibinfo {author} {\bibfnamefont {N.~R.}\
  \bibnamefont {Lobanov}}, \bibinfo {author} {\bibfnamefont {B.}~\bibnamefont
  {Lommel}}, \bibinfo {author} {\bibfnamefont {C.}~\bibnamefont {Mokry}},
  \bibinfo {author} {\bibfnamefont {E.}~\bibnamefont {Prasad}}, \bibinfo
  {author} {\bibfnamefont {D.~C.}\ \bibnamefont {Rafferty}}, \bibinfo {author}
  {\bibfnamefont {J.}~\bibnamefont {Runke}}, \bibinfo {author} {\bibfnamefont
  {K.}~\bibnamefont {Sekizawa}}, \bibinfo {author} {\bibfnamefont
  {C.}~\bibnamefont {Sengupta}}, \bibinfo {author} {\bibfnamefont
  {C.}~\bibnamefont {Simenel}}, \bibinfo {author} {\bibfnamefont {E.~C.}\
  \bibnamefont {Simpson}}, \bibinfo {author} {\bibfnamefont {J.~F.}\
  \bibnamefont {Smith}}, \bibinfo {author} {\bibfnamefont {P.}~\bibnamefont
  {Th\"{o}rle-Pospiech}}, \bibinfo {author} {\bibfnamefont {N.}~\bibnamefont
  {Trautmann}}, \bibinfo {author} {\bibfnamefont {K.}~\bibnamefont {Vo-Phuoc}},
  \bibinfo {author} {\bibfnamefont {J.}~\bibnamefont {Walshe}}, \bibinfo
  {author} {\bibfnamefont {E.}~\bibnamefont {Williams}},\ and\ \bibinfo
  {author} {\bibfnamefont {A.}~\bibnamefont {Yakushev}},\ }\bibfield  {title}
  {\bibinfo {title} {Zeptosecond contact times for element ${Z}=120$
  synthesis},\ }\href {https://doi.org/10.1016/j.physletb.2020.135626}
  {\bibfield  {journal} {\bibinfo  {journal} {Phys. Lett. B}\ }\textbf
  {\bibinfo {volume} {808}},\ \bibinfo {pages} {135626} (\bibinfo {year}
  {2020})}\BibitemShut {NoStop}%
\bibitem [{\citenamefont {Hagino}\ and\ \citenamefont
  {Takigawa}(2012)}]{Hagino2012_PTP128-1061}%
  \BibitemOpen
  \bibfield  {author} {\bibinfo {author} {\bibfnamefont {K.}~\bibnamefont
  {Hagino}}\ and\ \bibinfo {author} {\bibfnamefont {N.}~\bibnamefont
  {Takigawa}},\ }\bibfield  {title} {\bibinfo {title} {Subbarrier fusion
  reactions and many-particle quantum tunneling},\ }\href
  {https://doi.org/10.1143/PTP.128.1061} {\bibfield  {journal} {\bibinfo
  {journal} {Prog. Theo. Phys.}\ }\textbf {\bibinfo {volume} {128}},\ \bibinfo
  {pages} {1061} (\bibinfo {year} {2012})}\BibitemShut {NoStop}%
\bibitem [{\citenamefont {Back}\ \emph {et~al.}(2014)\citenamefont {Back},
  \citenamefont {Esbensen}, \citenamefont {Jiang},\ and\ \citenamefont
  {Rehm}}]{Back2014_RMP86-317360}%
  \BibitemOpen
  \bibfield  {author} {\bibinfo {author} {\bibfnamefont {B.~B.}\ \bibnamefont
  {Back}}, \bibinfo {author} {\bibfnamefont {H.}~\bibnamefont {Esbensen}},
  \bibinfo {author} {\bibfnamefont {C.~L.}\ \bibnamefont {Jiang}},\ and\
  \bibinfo {author} {\bibfnamefont {K.~E.}\ \bibnamefont {Rehm}},\ }\bibfield
  {title} {\bibinfo {title} {Recent developments in heavy-ion fusion
  reactions},\ }\href {https://doi.org/10.1103/RevModPhys.86.317} {\bibfield
  {journal} {\bibinfo  {journal} {Rev. Mod. Phys.}\ }\textbf {\bibinfo {volume}
  {86}},\ \bibinfo {pages} {317} (\bibinfo {year} {2014})}\BibitemShut
  {NoStop}%
\bibitem [{\citenamefont {\ifmmode~\acute{S}\else \'{S}\fi{}wi\k{a}tecki}\
  \emph {et~al.}(2005)\citenamefont {\ifmmode~\acute{S}\else
  \'{S}\fi{}wi\k{a}tecki}, \citenamefont {Siwek-Wilczy\ifmmode~\acute{n}\else
  \'{n}\fi{}ska},\ and\ \citenamefont {Wilczy\ifmmode~\acute{n}\else
  \'{n}\fi{}ski}}]{Swiatecki2005_PRC71-014602}%
  \BibitemOpen
  \bibfield  {author} {\bibinfo {author} {\bibfnamefont {W.~J.}\ \bibnamefont
  {\ifmmode~\acute{S}\else \'{S}\fi{}wi\k{a}tecki}}, \bibinfo {author}
  {\bibfnamefont {K.}~\bibnamefont {Siwek-Wilczy\ifmmode~\acute{n}\else
  \'{n}\fi{}ska}},\ and\ \bibinfo {author} {\bibfnamefont {J.}~\bibnamefont
  {Wilczy\ifmmode~\acute{n}\else \'{n}\fi{}ski}},\ }\bibfield  {title}
  {\bibinfo {title} {Fusion by diffusion. {I}{I}. synthesis of transfermium
  elements in cold fusion reactions},\ }\href
  {https://doi.org/10.1103/PhysRevC.71.014602} {\bibfield  {journal} {\bibinfo
  {journal} {Phys. Rev. C}\ }\textbf {\bibinfo {volume} {71}},\ \bibinfo
  {pages} {014602} (\bibinfo {year} {2005})}\BibitemShut {NoStop}%
\bibitem [{\citenamefont {Zagrebaev}\ and\ \citenamefont
  {Greiner}(2015)}]{Zagrebaev2015_NPA944-257}%
  \BibitemOpen
  \bibfield  {author} {\bibinfo {author} {\bibfnamefont {V.~I.}\ \bibnamefont
  {Zagrebaev}}\ and\ \bibinfo {author} {\bibfnamefont {W.}~\bibnamefont
  {Greiner}},\ }\bibfield  {title} {\bibinfo {title} {{Cross sections for the
  production of superheavy nuclei}},\ }\href
  {https://doi.org/10.1016/j.nuclphysa.2015.02.010} {\bibfield  {journal}
  {\bibinfo  {journal} {Nucl. Phys. A}\ }\textbf {\bibinfo {volume} {944}},\
  \bibinfo {pages} {257} (\bibinfo {year} {2015})}\BibitemShut {NoStop}%
\bibitem [{\citenamefont {Wang}\ \emph {et~al.}(2017)\citenamefont {Wang},
  \citenamefont {Wen}, \citenamefont {Zhao}, \citenamefont {Zhao},\ and\
  \citenamefont {Zhou}}]{Wang2017_ADNDT114-281370}%
  \BibitemOpen
  \bibfield  {author} {\bibinfo {author} {\bibfnamefont {B.}~\bibnamefont
  {Wang}}, \bibinfo {author} {\bibfnamefont {K.}~\bibnamefont {Wen}}, \bibinfo
  {author} {\bibfnamefont {W.-J.}\ \bibnamefont {Zhao}}, \bibinfo {author}
  {\bibfnamefont {E.-G.}\ \bibnamefont {Zhao}},\ and\ \bibinfo {author}
  {\bibfnamefont {S.-G.}\ \bibnamefont {Zhou}},\ }\bibfield  {title} {\bibinfo
  {title} {Systematics of capture and fusion dynamics in heavy-ion
  collisions},\ }\href {https://doi.org/10.1016/j.adt.2016.06.003} {\bibfield
  {journal} {\bibinfo  {journal} {At. Data Nucl. Data Tables}\ }\textbf
  {\bibinfo {volume} {114}},\ \bibinfo {pages} {281} (\bibinfo {year}
  {2017})}\BibitemShut {NoStop}%
\bibitem [{\citenamefont {Clerc}\ \emph {et~al.}(1984)\citenamefont {Clerc},
  \citenamefont {Keller}, \citenamefont {Sahm}, \citenamefont {Schmidt},
  \citenamefont {Schulte},\ and\ \citenamefont
  {Vermeulen}}]{Clerc1984_NPA419-571}%
  \BibitemOpen
  \bibfield  {author} {\bibinfo {author} {\bibfnamefont {H.-G.}\ \bibnamefont
  {Clerc}}, \bibinfo {author} {\bibfnamefont {J.~G.}\ \bibnamefont {Keller}},
  \bibinfo {author} {\bibfnamefont {C.-C.}\ \bibnamefont {Sahm}}, \bibinfo
  {author} {\bibfnamefont {K.-H.}\ \bibnamefont {Schmidt}}, \bibinfo {author}
  {\bibfnamefont {H.}~\bibnamefont {Schulte}},\ and\ \bibinfo {author}
  {\bibfnamefont {D.}~\bibnamefont {Vermeulen}},\ }\bibfield  {title} {\bibinfo
  {title} {{Fusion-fission and neutron-evaporation-residue cross-sections in
  $^{40}${A}r- and $^{50}${T}i-induced fusion reactions}},\ }\href
  {https://doi.org/10.1016/0375-9474(84)90634-1} {\bibfield  {journal}
  {\bibinfo  {journal} {Nucl. Phys. A}\ }\textbf {\bibinfo {volume} {419}},\
  \bibinfo {pages} {571} (\bibinfo {year} {1984})}\BibitemShut {NoStop}%
\bibitem [{\citenamefont {Pacheco}\ \emph {et~al.}(1992)\citenamefont
  {Pacheco}, \citenamefont {Fern\'andez~Niello}, \citenamefont {DiGregorio},
  \citenamefont {di~Tada}, \citenamefont {Testoni}, \citenamefont {Chan},
  \citenamefont {Ch\'avez}, \citenamefont {Gazes}, \citenamefont {Plagnol},\
  and\ \citenamefont {Stokstad}}]{Pacheco1992_PRC45-28612}%
  \BibitemOpen
  \bibfield  {author} {\bibinfo {author} {\bibfnamefont {A.~J.}\ \bibnamefont
  {Pacheco}}, \bibinfo {author} {\bibfnamefont {J.~O.}\ \bibnamefont
  {Fern\'andez~Niello}}, \bibinfo {author} {\bibfnamefont {D.~E.}\ \bibnamefont
  {DiGregorio}}, \bibinfo {author} {\bibfnamefont {M.}~\bibnamefont {di~Tada}},
  \bibinfo {author} {\bibfnamefont {J.~E.}\ \bibnamefont {Testoni}}, \bibinfo
  {author} {\bibfnamefont {Y.}~\bibnamefont {Chan}}, \bibinfo {author}
  {\bibfnamefont {E.}~\bibnamefont {Ch\'avez}}, \bibinfo {author}
  {\bibfnamefont {S.}~\bibnamefont {Gazes}}, \bibinfo {author} {\bibfnamefont
  {E.}~\bibnamefont {Plagnol}},\ and\ \bibinfo {author} {\bibfnamefont {R.~G.}\
  \bibnamefont {Stokstad}},\ }\bibfield  {title} {\bibinfo {title} {Capture
  reactions in the $^{40,48}\mathrm{Ca}$${+}^{197}${A}u and
  $^{40,48}\mathrm{Ca}$${+}^{208}${P}b systems},\ }\href
  {https://doi.org/10.1103/PhysRevC.45.2861} {\bibfield  {journal} {\bibinfo
  {journal} {Phys. Rev. C}\ }\textbf {\bibinfo {volume} {45}},\ \bibinfo
  {pages} {2861} (\bibinfo {year} {1992})}\BibitemShut {NoStop}%
\bibitem [{\citenamefont {Prokhorova}\ \emph {et~al.}(2008)\citenamefont
  {Prokhorova}, \citenamefont {Bogachev}, \citenamefont {Itkis}, \citenamefont
  {Itkis}, \citenamefont {Knyazheva}, \citenamefont {Kondratiev}, \citenamefont
  {Kozulin}, \citenamefont {Krupa}, \citenamefont {Oganessian}, \citenamefont
  {Pokrovsky}, \citenamefont {Pashkevich},\ and\ \citenamefont
  {Rusanov}}]{Prokhorova2008_NPA802-45}%
  \BibitemOpen
  \bibfield  {author} {\bibinfo {author} {\bibfnamefont {E.}~\bibnamefont
  {Prokhorova}}, \bibinfo {author} {\bibfnamefont {A.}~\bibnamefont
  {Bogachev}}, \bibinfo {author} {\bibfnamefont {M.}~\bibnamefont {Itkis}},
  \bibinfo {author} {\bibfnamefont {I.}~\bibnamefont {Itkis}}, \bibinfo
  {author} {\bibfnamefont {G.}~\bibnamefont {Knyazheva}}, \bibinfo {author}
  {\bibfnamefont {N.}~\bibnamefont {Kondratiev}}, \bibinfo {author}
  {\bibfnamefont {E.}~\bibnamefont {Kozulin}}, \bibinfo {author} {\bibfnamefont
  {L.}~\bibnamefont {Krupa}}, \bibinfo {author} {\bibfnamefont
  {Y.}~\bibnamefont {Oganessian}}, \bibinfo {author} {\bibfnamefont
  {I.}~\bibnamefont {Pokrovsky}}, \bibinfo {author} {\bibfnamefont
  {V.}~\bibnamefont {Pashkevich}},\ and\ \bibinfo {author} {\bibfnamefont
  {A.}~\bibnamefont {Rusanov}},\ }\bibfield  {title} {\bibinfo {title} {The
  fusion--fission and quasi--fission processes in the reaction $^{48}${C}a +
  $^{208}${P}b at energies near the coulomb barrier},\ }\href
  {https://doi.org/https://doi.org/10.1016/j.nuclphysa.2008.01.016} {\bibfield
  {journal} {\bibinfo  {journal} {Nucl. Phys. A}\ }\textbf {\bibinfo {volume}
  {802}},\ \bibinfo {pages} {45} (\bibinfo {year} {2008})}\BibitemShut
  {NoStop}%
\bibitem [{\citenamefont {Banerjee}\ \emph {et~al.}(2019)\citenamefont
  {Banerjee}, \citenamefont {Hinde}, \citenamefont {Dasgupta}, \citenamefont
  {Simpson}, \citenamefont {Jeung}, \citenamefont {Simenel}, \citenamefont
  {Swinton-Bland}, \citenamefont {Williams}, \citenamefont {Carter},
  \citenamefont {Cook}, \citenamefont {David}, \citenamefont {D\"ullmann},
  \citenamefont {Khuyagbaatar}, \citenamefont {Kindler}, \citenamefont
  {Lommel}, \citenamefont {Prasad}, \citenamefont {Sengupta}, \citenamefont
  {Smith}, \citenamefont {Vo-Phuoc}, \citenamefont {Walshe},\ and\
  \citenamefont {Yakushev}}]{Banerjee2019_PRL122-232503}%
  \BibitemOpen
  \bibfield  {author} {\bibinfo {author} {\bibfnamefont {K.}~\bibnamefont
  {Banerjee}}, \bibinfo {author} {\bibfnamefont {D.~J.}\ \bibnamefont {Hinde}},
  \bibinfo {author} {\bibfnamefont {M.}~\bibnamefont {Dasgupta}}, \bibinfo
  {author} {\bibfnamefont {E.~C.}\ \bibnamefont {Simpson}}, \bibinfo {author}
  {\bibfnamefont {D.~Y.}\ \bibnamefont {Jeung}}, \bibinfo {author}
  {\bibfnamefont {C.}~\bibnamefont {Simenel}}, \bibinfo {author} {\bibfnamefont
  {B.~M.~A.}\ \bibnamefont {Swinton-Bland}}, \bibinfo {author} {\bibfnamefont
  {E.}~\bibnamefont {Williams}}, \bibinfo {author} {\bibfnamefont {I.~P.}\
  \bibnamefont {Carter}}, \bibinfo {author} {\bibfnamefont {K.~J.}\
  \bibnamefont {Cook}}, \bibinfo {author} {\bibfnamefont {H.~M.}\ \bibnamefont
  {David}}, \bibinfo {author} {\bibfnamefont {C.~E.}\ \bibnamefont
  {D\"ullmann}}, \bibinfo {author} {\bibfnamefont {J.}~\bibnamefont
  {Khuyagbaatar}}, \bibinfo {author} {\bibfnamefont {B.}~\bibnamefont
  {Kindler}}, \bibinfo {author} {\bibfnamefont {B.}~\bibnamefont {Lommel}},
  \bibinfo {author} {\bibfnamefont {E.}~\bibnamefont {Prasad}}, \bibinfo
  {author} {\bibfnamefont {C.}~\bibnamefont {Sengupta}}, \bibinfo {author}
  {\bibfnamefont {J.~F.}\ \bibnamefont {Smith}}, \bibinfo {author}
  {\bibfnamefont {K.}~\bibnamefont {Vo-Phuoc}}, \bibinfo {author}
  {\bibfnamefont {J.}~\bibnamefont {Walshe}},\ and\ \bibinfo {author}
  {\bibfnamefont {A.}~\bibnamefont {Yakushev}},\ }\bibfield  {title} {\bibinfo
  {title} {Mechanisms suppressing superheavy element yields in cold fusion
  reactions},\ }\href {https://doi.org/10.1103/PhysRevLett.122.232503}
  {\bibfield  {journal} {\bibinfo  {journal} {Phys. Rev. Lett.}\ }\textbf
  {\bibinfo {volume} {122}},\ \bibinfo {pages} {232503} (\bibinfo {year}
  {2019})}\BibitemShut {NoStop}%
\bibitem [{\citenamefont {Adamian}\ \emph {et~al.}(2000)\citenamefont
  {Adamian}, \citenamefont {Antonenko}, \citenamefont {Ivanova},\ and\
  \citenamefont {Scheid}}]{Adamian2000_PRC62-064303}%
  \BibitemOpen
  \bibfield  {author} {\bibinfo {author} {\bibfnamefont {G.~G.}\ \bibnamefont
  {Adamian}}, \bibinfo {author} {\bibfnamefont {N.~V.}\ \bibnamefont
  {Antonenko}}, \bibinfo {author} {\bibfnamefont {S.~P.}\ \bibnamefont
  {Ivanova}},\ and\ \bibinfo {author} {\bibfnamefont {W.}~\bibnamefont
  {Scheid}},\ }\bibfield  {title} {\bibinfo {title} {Analysis of survival
  probability of superheavy nuclei},\ }\href
  {https://doi.org/10.1103/PhysRevC.62.064303} {\bibfield  {journal} {\bibinfo
  {journal} {Phys. Rev. C}\ }\textbf {\bibinfo {volume} {62}},\ \bibinfo
  {pages} {064303} (\bibinfo {year} {2000})}\BibitemShut {NoStop}%
\bibitem [{\citenamefont {Antonenko}\ \emph {et~al.}(1993)\citenamefont
  {Antonenko}, \citenamefont {Cherepanov}, \citenamefont {Nasirov},
  \citenamefont {Permjakov},\ and\ \citenamefont
  {Volkov}}]{Antonenko1993_PLB319-425}%
  \BibitemOpen
  \bibfield  {author} {\bibinfo {author} {\bibfnamefont {N.~V.}\ \bibnamefont
  {Antonenko}}, \bibinfo {author} {\bibfnamefont {E.~A.}\ \bibnamefont
  {Cherepanov}}, \bibinfo {author} {\bibfnamefont {A.~K.}\ \bibnamefont
  {Nasirov}}, \bibinfo {author} {\bibfnamefont {V.~P.}\ \bibnamefont
  {Permjakov}},\ and\ \bibinfo {author} {\bibfnamefont {V.~V.}\ \bibnamefont
  {Volkov}},\ }\bibfield  {title} {\bibinfo {title} {{Competition between
  complete fusion and quasi-fission in reactions between massive nuclei. The
  fusion barrier}},\ }\href {https://doi.org/10.1016/0370-2693(93)91746-A}
  {\bibfield  {journal} {\bibinfo  {journal} {Phys. Lett. B}\ }\textbf
  {\bibinfo {volume} {319}},\ \bibinfo {pages} {425} (\bibinfo {year}
  {1993})}\BibitemShut {NoStop}%
\bibitem [{\citenamefont {Adamian}\ \emph {et~al.}(1997)\citenamefont
  {Adamian}, \citenamefont {Antonenko}, \citenamefont {Scheid},\ and\
  \citenamefont {Volkov}}]{Adamian1997_NPA627-361}%
  \BibitemOpen
  \bibfield  {author} {\bibinfo {author} {\bibfnamefont {G.~G.}\ \bibnamefont
  {Adamian}}, \bibinfo {author} {\bibfnamefont {N.~V.}\ \bibnamefont
  {Antonenko}}, \bibinfo {author} {\bibfnamefont {W.}~\bibnamefont {Scheid}},\
  and\ \bibinfo {author} {\bibfnamefont {V.~V.}\ \bibnamefont {Volkov}},\
  }\bibfield  {title} {\bibinfo {title} {Treatment of competition between
  complete fusion and quasifission in collisions of heavy nuclei},\ }\href
  {https://doi.org/10.1016/S0375-9474(97)00605-2} {\bibfield  {journal}
  {\bibinfo  {journal} {Nucl. Phys. A}\ }\textbf {\bibinfo {volume} {627}},\
  \bibinfo {pages} {361} (\bibinfo {year} {1997})}\BibitemShut {NoStop}%
\bibitem [{\citenamefont {Feng}\ \emph {et~al.}(2006)\citenamefont {Feng},
  \citenamefont {Jin}, \citenamefont {Fu},\ and\ \citenamefont
  {Li}}]{Feng2006_NPA771-50}%
  \BibitemOpen
  \bibfield  {author} {\bibinfo {author} {\bibfnamefont {Z.-Q.}\ \bibnamefont
  {Feng}}, \bibinfo {author} {\bibfnamefont {G.-M.}\ \bibnamefont {Jin}},
  \bibinfo {author} {\bibfnamefont {F.}~\bibnamefont {Fu}},\ and\ \bibinfo
  {author} {\bibfnamefont {J.-Q.}\ \bibnamefont {Li}},\ }\bibfield  {title}
  {\bibinfo {title} {Production cross sections of superheavy nuclei based on
  dinuclear system model},\ }\href
  {https://doi.org/10.1016/j.nuclphysa.2006.03.002} {\bibfield  {journal}
  {\bibinfo  {journal} {Nucl. Phys. A}\ }\textbf {\bibinfo {volume} {771}},\
  \bibinfo {pages} {50} (\bibinfo {year} {2006})}\BibitemShut {NoStop}%
\bibitem [{\citenamefont {Zhu}\ \emph {et~al.}(2014)\citenamefont {Zhu},
  \citenamefont {Xie},\ and\ \citenamefont {Zhang}}]{Zhu2014_PRC89-024615}%
  \BibitemOpen
  \bibfield  {author} {\bibinfo {author} {\bibfnamefont {L.}~\bibnamefont
  {Zhu}}, \bibinfo {author} {\bibfnamefont {W.-J.}\ \bibnamefont {Xie}},\ and\
  \bibinfo {author} {\bibfnamefont {F.-S.}\ \bibnamefont {Zhang}},\ }\bibfield
  {title} {\bibinfo {title} {Production cross sections of superheavy elements
  ${Z}=119$ and 120 in hot fusion reactions},\ }\href
  {https://doi.org/10.1103/PhysRevC.89.024615} {\bibfield  {journal} {\bibinfo
  {journal} {Phys. Rev. C}\ }\textbf {\bibinfo {volume} {89}},\ \bibinfo
  {pages} {024615} (\bibinfo {year} {2014})}\BibitemShut {NoStop}%
\bibitem [{\citenamefont {Wang}\ \emph {et~al.}(2012)\citenamefont {Wang},
  \citenamefont {Zhao}, \citenamefont {Scheid},\ and\ \citenamefont
  {Zhou}}]{Wang2012_PRC85-041601R}%
  \BibitemOpen
  \bibfield  {author} {\bibinfo {author} {\bibfnamefont {N.}~\bibnamefont
  {Wang}}, \bibinfo {author} {\bibfnamefont {E.-G.}\ \bibnamefont {Zhao}},
  \bibinfo {author} {\bibfnamefont {W.}~\bibnamefont {Scheid}},\ and\ \bibinfo
  {author} {\bibfnamefont {S.-G.}\ \bibnamefont {Zhou}},\ }\bibfield  {title}
  {\bibinfo {title} {Theoretical study of the synthesis of superheavy nuclei
  with ${Z}=119$ and 120 in heavy-ion reactions with trans-uranium targets},\
  }\href {https://doi.org/10.1103/PhysRevC.85.041601} {\bibfield  {journal}
  {\bibinfo  {journal} {Phys. Rev. C}\ }\textbf {\bibinfo {volume} {85}},\
  \bibinfo {pages} {041601(R)} (\bibinfo {year} {2012})}\BibitemShut {NoStop}%
\bibitem [{\citenamefont {Liu}\ and\ \citenamefont
  {Bao}(2011)}]{Liu2011_PRC84-031602R}%
  \BibitemOpen
  \bibfield  {author} {\bibinfo {author} {\bibfnamefont {Z.-H.}\ \bibnamefont
  {Liu}}\ and\ \bibinfo {author} {\bibfnamefont {J.-D.}\ \bibnamefont {Bao}},\
  }\bibfield  {title} {\bibinfo {title} {Calculation of the evaporation residue
  cross sections for the synthesis of the superheavy element ${Z}=119$ via the
  ${}^{50}${T}i+${}^{249}${B}k hot fusion reaction},\ }\href
  {https://doi.org/10.1103/PhysRevC.84.031602} {\bibfield  {journal} {\bibinfo
  {journal} {Phys. Rev. C}\ }\textbf {\bibinfo {volume} {84}},\ \bibinfo
  {pages} {031602} (\bibinfo {year} {2011})}\BibitemShut {NoStop}%
\bibitem [{\citenamefont {Cap}\ \emph {et~al.}(2011)\citenamefont {Cap},
  \citenamefont {Siwek-Wilczy\ifmmode~\acute{n}\else \'{n}\fi{}ska},\ and\
  \citenamefont {Wilczy\ifmmode~\acute{n}\else
  \'{n}\fi{}ski}}]{Cap2011_PRC83-054602}%
  \BibitemOpen
  \bibfield  {author} {\bibinfo {author} {\bibfnamefont {T.}~\bibnamefont
  {Cap}}, \bibinfo {author} {\bibfnamefont {K.}~\bibnamefont
  {Siwek-Wilczy\ifmmode~\acute{n}\else \'{n}\fi{}ska}},\ and\ \bibinfo {author}
  {\bibfnamefont {J.}~\bibnamefont {Wilczy\ifmmode~\acute{n}\else
  \'{n}\fi{}ski}},\ }\bibfield  {title} {\bibinfo {title} {Nucleus-nucleus cold
  fusion reactions analyzed with the $l$-dependent ``fusion by diffusion''
  model},\ }\href {https://doi.org/10.1103/PhysRevC.83.054602} {\bibfield
  {journal} {\bibinfo  {journal} {Phys. Rev. C}\ }\textbf {\bibinfo {volume}
  {83}},\ \bibinfo {pages} {054602} (\bibinfo {year} {2011})}\BibitemShut
  {NoStop}%
\bibitem [{\citenamefont {Cap}\ \emph {et~al.}(2013)\citenamefont {Cap},
  \citenamefont {Siwek-Wilczy\ifmmode~\acute{n}\else \'{n}\fi{}ska},
  \citenamefont {Kowal},\ and\ \citenamefont {Wilczy\ifmmode~\acute{n}\else
  \'{n}\fi{}ski}}]{Cap2013_PRC88-037603}%
  \BibitemOpen
  \bibfield  {author} {\bibinfo {author} {\bibfnamefont {T.}~\bibnamefont
  {Cap}}, \bibinfo {author} {\bibfnamefont {K.}~\bibnamefont
  {Siwek-Wilczy\ifmmode~\acute{n}\else \'{n}\fi{}ska}}, \bibinfo {author}
  {\bibfnamefont {M.}~\bibnamefont {Kowal}},\ and\ \bibinfo {author}
  {\bibfnamefont {J.}~\bibnamefont {Wilczy\ifmmode~\acute{n}\else
  \'{n}\fi{}ski}},\ }\bibfield  {title} {\bibinfo {title} {Calculations of the
  cross sections for the synthesis of new ${}^{293-296}$118 isotopes in
  ${}^{249-252}${C}f(${}^{48}${C}a,$xn$) reactions},\ }\href
  {https://doi.org/10.1103/PhysRevC.88.037603} {\bibfield  {journal} {\bibinfo
  {journal} {Phys. Rev. C}\ }\textbf {\bibinfo {volume} {88}},\ \bibinfo
  {pages} {037603} (\bibinfo {year} {2013})}\BibitemShut {NoStop}%
\bibitem [{\citenamefont {Bao}\ \emph {et~al.}(2017)\citenamefont {Bao},
  \citenamefont {Guo}, \citenamefont {Zhang},\ and\ \citenamefont
  {Li}}]{Bao2017_PRC96-024610}%
  \BibitemOpen
  \bibfield  {author} {\bibinfo {author} {\bibfnamefont {X.~J.}\ \bibnamefont
  {Bao}}, \bibinfo {author} {\bibfnamefont {S.~Q.}\ \bibnamefont {Guo}},
  \bibinfo {author} {\bibfnamefont {H.~F.}\ \bibnamefont {Zhang}},\ and\
  \bibinfo {author} {\bibfnamefont {J.~Q.}\ \bibnamefont {Li}},\ }\bibfield
  {title} {\bibinfo {title} {Influence of entrance channel on production cross
  sections of superheavy nuclei},\ }\href
  {https://doi.org/10.1103/PhysRevC.96.024610} {\bibfield  {journal} {\bibinfo
  {journal} {Phys. Rev. C}\ }\textbf {\bibinfo {volume} {96}},\ \bibinfo
  {pages} {024610} (\bibinfo {year} {2017})}\BibitemShut {NoStop}%
\bibitem [{\citenamefont {Hagino}(2018)}]{Hagino2018_PRC98-014607}%
  \BibitemOpen
  \bibfield  {author} {\bibinfo {author} {\bibfnamefont {K.}~\bibnamefont
  {Hagino}},\ }\bibfield  {title} {\bibinfo {title} {Hot fusion reactions with
  deformed nuclei for synthesis of superheavy nuclei: An extension of the
  fusion-by-diffusion model},\ }\href
  {https://doi.org/10.1103/PhysRevC.98.014607} {\bibfield  {journal} {\bibinfo
  {journal} {Phys. Rev. C}\ }\textbf {\bibinfo {volume} {98}},\ \bibinfo
  {pages} {014607} (\bibinfo {year} {2018})}\BibitemShut {NoStop}%
\bibitem [{\citenamefont {Lv}\ \emph {et~al.}(2021)\citenamefont {Lv},
  \citenamefont {Yue}, \citenamefont {Zhao},\ and\ \citenamefont
  {Wang}}]{Lv2021_PRC103-064616}%
  \BibitemOpen
  \bibfield  {author} {\bibinfo {author} {\bibfnamefont {X.-J.}\ \bibnamefont
  {Lv}}, \bibinfo {author} {\bibfnamefont {Z.-Y.}\ \bibnamefont {Yue}},
  \bibinfo {author} {\bibfnamefont {W.-J.}\ \bibnamefont {Zhao}},\ and\
  \bibinfo {author} {\bibfnamefont {B.}~\bibnamefont {Wang}},\ }\bibfield
  {title} {\bibinfo {title} {Theoretical study of evaporation-residue cross
  sections of superheavy nuclei},\ }\href
  {https://doi.org/10.1103/PhysRevC.103.064616} {\bibfield  {journal} {\bibinfo
   {journal} {Phys. Rev. C}\ }\textbf {\bibinfo {volume} {103}},\ \bibinfo
  {pages} {064616} (\bibinfo {year} {2021})}\BibitemShut {NoStop}%
\bibitem [{\citenamefont {Loveland}(2007)}]{Loveland2007_PRC76-014612}%
  \BibitemOpen
  \bibfield  {author} {\bibinfo {author} {\bibfnamefont {W.}~\bibnamefont
  {Loveland}},\ }\bibfield  {title} {\bibinfo {title} {Synthesis of
  transactinide nuclei using radioactive beams},\ }\href
  {https://doi.org/10.1103/PhysRevC.76.014612} {\bibfield  {journal} {\bibinfo
  {journal} {Phys. Rev. C}\ }\textbf {\bibinfo {volume} {76}},\ \bibinfo
  {pages} {014612} (\bibinfo {year} {2007})}\BibitemShut {NoStop}%
\bibitem [{\citenamefont {Naik}\ \emph {et~al.}(2007)\citenamefont {Naik},
  \citenamefont {Loveland}, \citenamefont {Sprunger}, \citenamefont
  {Vinodkumar}, \citenamefont {Peterson}, \citenamefont {Jiang}, \citenamefont
  {Zhu}, \citenamefont {Tang}, \citenamefont {Moore},\ and\ \citenamefont
  {Chowdhury}}]{Naik2007_PRC76-054604}%
  \BibitemOpen
  \bibfield  {author} {\bibinfo {author} {\bibfnamefont {R.~S.}\ \bibnamefont
  {Naik}}, \bibinfo {author} {\bibfnamefont {W.}~\bibnamefont {Loveland}},
  \bibinfo {author} {\bibfnamefont {P.~H.}\ \bibnamefont {Sprunger}}, \bibinfo
  {author} {\bibfnamefont {A.~M.}\ \bibnamefont {Vinodkumar}}, \bibinfo
  {author} {\bibfnamefont {D.}~\bibnamefont {Peterson}}, \bibinfo {author}
  {\bibfnamefont {C.~L.}\ \bibnamefont {Jiang}}, \bibinfo {author}
  {\bibfnamefont {S.}~\bibnamefont {Zhu}}, \bibinfo {author} {\bibfnamefont
  {X.}~\bibnamefont {Tang}}, \bibinfo {author} {\bibfnamefont {E.~F.}\
  \bibnamefont {Moore}},\ and\ \bibinfo {author} {\bibfnamefont
  {P.}~\bibnamefont {Chowdhury}},\ }\bibfield  {title} {\bibinfo {title}
  {Measurement of the fusion probability ${P}_{\mathrm{{c}{n}}}$ for the
  reaction of $^{50}\mathrm{Ti}$ with $^{208}\mathrm{Pb}$},\ }\href
  {https://doi.org/10.1103/PhysRevC.76.054604} {\bibfield  {journal} {\bibinfo
  {journal} {Phys. Rev. C}\ }\textbf {\bibinfo {volume} {76}},\ \bibinfo
  {pages} {054604} (\bibinfo {year} {2007})}\BibitemShut {NoStop}%
\bibitem [{\citenamefont {Yanez}\ \emph {et~al.}(2013)\citenamefont {Yanez},
  \citenamefont {Loveland}, \citenamefont {Barrett}, \citenamefont {Yao},
  \citenamefont {Back}, \citenamefont {Zhu},\ and\ \citenamefont
  {Khoo}}]{Yanez2013_PRC88-014606}%
  \BibitemOpen
  \bibfield  {author} {\bibinfo {author} {\bibfnamefont {R.}~\bibnamefont
  {Yanez}}, \bibinfo {author} {\bibfnamefont {W.}~\bibnamefont {Loveland}},
  \bibinfo {author} {\bibfnamefont {J.~S.}\ \bibnamefont {Barrett}}, \bibinfo
  {author} {\bibfnamefont {L.}~\bibnamefont {Yao}}, \bibinfo {author}
  {\bibfnamefont {B.~B.}\ \bibnamefont {Back}}, \bibinfo {author}
  {\bibfnamefont {S.}~\bibnamefont {Zhu}},\ and\ \bibinfo {author}
  {\bibfnamefont {T.~L.}\ \bibnamefont {Khoo}},\ }\bibfield  {title} {\bibinfo
  {title} {Measurement of the fusion probability, ${P}_{\mathrm{{c}{n}}}$, for
  hot fusion reactions},\ }\href {https://doi.org/10.1103/PhysRevC.88.014606}
  {\bibfield  {journal} {\bibinfo  {journal} {Phys. Rev. C}\ }\textbf {\bibinfo
  {volume} {88}},\ \bibinfo {pages} {014606} (\bibinfo {year}
  {2013})}\BibitemShut {NoStop}%
\bibitem [{\citenamefont {L\"u}\ \emph {et~al.}(2016)\citenamefont {L\"u},
  \citenamefont {Boilley}, \citenamefont {Abe},\ and\ \citenamefont
  {Shen}}]{Lu2016_PRC94-034616}%
  \BibitemOpen
  \bibfield  {author} {\bibinfo {author} {\bibfnamefont {H.}~\bibnamefont
  {L\"u}}, \bibinfo {author} {\bibfnamefont {D.}~\bibnamefont {Boilley}},
  \bibinfo {author} {\bibfnamefont {Y.}~\bibnamefont {Abe}},\ and\ \bibinfo
  {author} {\bibfnamefont {C.}~\bibnamefont {Shen}},\ }\bibfield  {title}
  {\bibinfo {title} {Synthesis of superheavy elements: Uncertainty analysis to
  improve the predictive power of reaction models},\ }\href
  {https://doi.org/10.1103/PhysRevC.94.034616} {\bibfield  {journal} {\bibinfo
  {journal} {Phys. Rev. C}\ }\textbf {\bibinfo {volume} {94}},\ \bibinfo
  {pages} {034616} (\bibinfo {year} {2016})}\BibitemShut {NoStop}%
\bibitem [{\citenamefont {Zagrebaev}\ and\ \citenamefont
  {Greiner}(2008)}]{Zagrebaev2008_PRC78-034610}%
  \BibitemOpen
  \bibfield  {author} {\bibinfo {author} {\bibfnamefont {V.}~\bibnamefont
  {Zagrebaev}}\ and\ \bibinfo {author} {\bibfnamefont {W.}~\bibnamefont
  {Greiner}},\ }\bibfield  {title} {\bibinfo {title} {Synthesis of superheavy
  nuclei: A search for new production reactions},\ }\href
  {https://doi.org/10.1103/PhysRevC.78.034610} {\bibfield  {journal} {\bibinfo
  {journal} {Phys. Rev. C}\ }\textbf {\bibinfo {volume} {78}},\ \bibinfo
  {pages} {034610} (\bibinfo {year} {2008})}\BibitemShut {NoStop}%
\bibitem [{\citenamefont {Berriman}\ \emph {et~al.}(2001)\citenamefont
  {Berriman}, \citenamefont {Hinde}, \citenamefont {Dasgupta}, \citenamefont
  {Morton}, \citenamefont {Butt},\ and\ \citenamefont
  {Newton}}]{Berriman2001_Nature413-144}%
  \BibitemOpen
  \bibfield  {author} {\bibinfo {author} {\bibfnamefont {A.~C.}\ \bibnamefont
  {Berriman}}, \bibinfo {author} {\bibfnamefont {D.~J.}\ \bibnamefont {Hinde}},
  \bibinfo {author} {\bibfnamefont {M.}~\bibnamefont {Dasgupta}}, \bibinfo
  {author} {\bibfnamefont {C.~R.}\ \bibnamefont {Morton}}, \bibinfo {author}
  {\bibfnamefont {R.~D.}\ \bibnamefont {Butt}},\ and\ \bibinfo {author}
  {\bibfnamefont {J.~O.}\ \bibnamefont {Newton}},\ }\bibfield  {title}
  {\bibinfo {title} {Unexpected inhibition of fusion in
  nucleus{\textendash}nucleus collisions},\ }\href
  {https://doi.org/10.1038/35093069} {\bibfield  {journal} {\bibinfo  {journal}
  {Nature}\ }\textbf {\bibinfo {volume} {413}},\ \bibinfo {pages} {144}
  (\bibinfo {year} {2001})}\BibitemShut {NoStop}%
\bibitem [{\citenamefont {Khuyagbaatar}\ \emph {et~al.}(2012)\citenamefont
  {Khuyagbaatar}, \citenamefont {Nishio}, \citenamefont {Hofmann},
  \citenamefont {Ackermann}, \citenamefont {Block}, \citenamefont {Heinz},
  \citenamefont {He\ss{}berger}, \citenamefont {Hirose}, \citenamefont
  {Ikezoe}, \citenamefont {Kinlder}, \citenamefont {Lommel}, \citenamefont
  {Makii}, \citenamefont {Mitsuoka}, \citenamefont {Nishinaka}, \citenamefont
  {Ohtsuki}, \citenamefont {Wakabayashi},\ and\ \citenamefont
  {Yan}}]{Khuyagbaatar2012_PRC86-064602}%
  \BibitemOpen
  \bibfield  {author} {\bibinfo {author} {\bibfnamefont {J.}~\bibnamefont
  {Khuyagbaatar}}, \bibinfo {author} {\bibfnamefont {K.}~\bibnamefont
  {Nishio}}, \bibinfo {author} {\bibfnamefont {S.}~\bibnamefont {Hofmann}},
  \bibinfo {author} {\bibfnamefont {D.}~\bibnamefont {Ackermann}}, \bibinfo
  {author} {\bibfnamefont {M.}~\bibnamefont {Block}}, \bibinfo {author}
  {\bibfnamefont {S.}~\bibnamefont {Heinz}}, \bibinfo {author} {\bibfnamefont
  {F.~P.}\ \bibnamefont {He\ss{}berger}}, \bibinfo {author} {\bibfnamefont
  {K.}~\bibnamefont {Hirose}}, \bibinfo {author} {\bibfnamefont
  {H.}~\bibnamefont {Ikezoe}}, \bibinfo {author} {\bibfnamefont
  {B.}~\bibnamefont {Kinlder}}, \bibinfo {author} {\bibfnamefont
  {B.}~\bibnamefont {Lommel}}, \bibinfo {author} {\bibfnamefont
  {H.}~\bibnamefont {Makii}}, \bibinfo {author} {\bibfnamefont
  {S.}~\bibnamefont {Mitsuoka}}, \bibinfo {author} {\bibfnamefont
  {I.}~\bibnamefont {Nishinaka}}, \bibinfo {author} {\bibfnamefont
  {T.}~\bibnamefont {Ohtsuki}}, \bibinfo {author} {\bibfnamefont
  {Y.}~\bibnamefont {Wakabayashi}},\ and\ \bibinfo {author} {\bibfnamefont
  {S.}~\bibnamefont {Yan}},\ }\bibfield  {title} {\bibinfo {title} {Evidence
  for hindrance in fusion between sulfur and lead nuclei},\ }\href
  {https://doi.org/10.1103/PhysRevC.86.064602} {\bibfield  {journal} {\bibinfo
  {journal} {Phys. Rev. C}\ }\textbf {\bibinfo {volume} {86}},\ \bibinfo
  {pages} {064602} (\bibinfo {year} {2012})}\BibitemShut {NoStop}%
\bibitem [{\citenamefont {Lin}\ \emph {et~al.}(2012)\citenamefont {Lin},
  \citenamefont {du~Rietz}, \citenamefont {Hinde}, \citenamefont {Dasgupta},
  \citenamefont {Thomas}, \citenamefont {Brown}, \citenamefont {Evers},
  \citenamefont {Gasques},\ and\ \citenamefont
  {Rodriguez}}]{Lin2012_PRC85-014611}%
  \BibitemOpen
  \bibfield  {author} {\bibinfo {author} {\bibfnamefont {C.~J.}\ \bibnamefont
  {Lin}}, \bibinfo {author} {\bibfnamefont {R.}~\bibnamefont {du~Rietz}},
  \bibinfo {author} {\bibfnamefont {D.~J.}\ \bibnamefont {Hinde}}, \bibinfo
  {author} {\bibfnamefont {M.}~\bibnamefont {Dasgupta}}, \bibinfo {author}
  {\bibfnamefont {R.~G.}\ \bibnamefont {Thomas}}, \bibinfo {author}
  {\bibfnamefont {M.~L.}\ \bibnamefont {Brown}}, \bibinfo {author}
  {\bibfnamefont {M.}~\bibnamefont {Evers}}, \bibinfo {author} {\bibfnamefont
  {L.~R.}\ \bibnamefont {Gasques}},\ and\ \bibinfo {author} {\bibfnamefont
  {M.~D.}\ \bibnamefont {Rodriguez}},\ }\bibfield  {title} {\bibinfo {title}
  {Systematic behavior of mass distributions in ${}^{48}${T}i-induced fission
  at near-barrier energies},\ }\href
  {https://doi.org/10.1103/PhysRevC.85.014611} {\bibfield  {journal} {\bibinfo
  {journal} {Phys. Rev. C}\ }\textbf {\bibinfo {volume} {85}},\ \bibinfo
  {pages} {014611} (\bibinfo {year} {2012})}\BibitemShut {NoStop}%
\bibitem [{\citenamefont {Hammerton}\ \emph {et~al.}(2015)\citenamefont
  {Hammerton}, \citenamefont {Kohley}, \citenamefont {Hinde}, \citenamefont
  {Dasgupta}, \citenamefont {Wakhle}, \citenamefont {Williams}, \citenamefont
  {Oberacker}, \citenamefont {Umar}, \citenamefont {Carter}, \citenamefont
  {Cook}, \citenamefont {Greene}, \citenamefont {Jeung}, \citenamefont {Luong},
  \citenamefont {McNeil}, \citenamefont {Palshetkar}, \citenamefont {Rafferty},
  \citenamefont {Simenel},\ and\ \citenamefont
  {Stiefel}}]{Hammerton2015_PRC91-041602}%
  \BibitemOpen
  \bibfield  {author} {\bibinfo {author} {\bibfnamefont {K.}~\bibnamefont
  {Hammerton}}, \bibinfo {author} {\bibfnamefont {Z.}~\bibnamefont {Kohley}},
  \bibinfo {author} {\bibfnamefont {D.~J.}\ \bibnamefont {Hinde}}, \bibinfo
  {author} {\bibfnamefont {M.}~\bibnamefont {Dasgupta}}, \bibinfo {author}
  {\bibfnamefont {A.}~\bibnamefont {Wakhle}}, \bibinfo {author} {\bibfnamefont
  {E.}~\bibnamefont {Williams}}, \bibinfo {author} {\bibfnamefont {V.~E.}\
  \bibnamefont {Oberacker}}, \bibinfo {author} {\bibfnamefont {A.~S.}\
  \bibnamefont {Umar}}, \bibinfo {author} {\bibfnamefont {I.~P.}\ \bibnamefont
  {Carter}}, \bibinfo {author} {\bibfnamefont {K.~J.}\ \bibnamefont {Cook}},
  \bibinfo {author} {\bibfnamefont {J.}~\bibnamefont {Greene}}, \bibinfo
  {author} {\bibfnamefont {D.~Y.}\ \bibnamefont {Jeung}}, \bibinfo {author}
  {\bibfnamefont {D.~H.}\ \bibnamefont {Luong}}, \bibinfo {author}
  {\bibfnamefont {S.~D.}\ \bibnamefont {McNeil}}, \bibinfo {author}
  {\bibfnamefont {C.~S.}\ \bibnamefont {Palshetkar}}, \bibinfo {author}
  {\bibfnamefont {D.~C.}\ \bibnamefont {Rafferty}}, \bibinfo {author}
  {\bibfnamefont {C.}~\bibnamefont {Simenel}},\ and\ \bibinfo {author}
  {\bibfnamefont {K.}~\bibnamefont {Stiefel}},\ }\bibfield  {title} {\bibinfo
  {title} {Reduced quasifission competition in fusion reactions forming
  neutron-rich heavy elements},\ }\href
  {https://doi.org/10.1103/PhysRevC.91.041602} {\bibfield  {journal} {\bibinfo
  {journal} {Phys. Rev. C}\ }\textbf {\bibinfo {volume} {91}},\ \bibinfo
  {pages} {041602} (\bibinfo {year} {2015})}\BibitemShut {NoStop}%
\bibitem [{\citenamefont {Back}(1985)}]{Back1985_PRC31-2104}%
  \BibitemOpen
  \bibfield  {author} {\bibinfo {author} {\bibfnamefont {B.~B.}\ \bibnamefont
  {Back}},\ }\bibfield  {title} {\bibinfo {title} {Complete fusion and
  quasifission in reactions between heavy ions},\ }\href
  {https://doi.org/10.1103/PhysRevC.31.2104} {\bibfield  {journal} {\bibinfo
  {journal} {Phys. Rev. C}\ }\textbf {\bibinfo {volume} {31}},\ \bibinfo
  {pages} {2104} (\bibinfo {year} {1985})}\BibitemShut {NoStop}%
\bibitem [{\citenamefont {Tsang}\ \emph {et~al.}(1983)\citenamefont {Tsang},
  \citenamefont {Utsunomiya}, \citenamefont {Gelbke}, \citenamefont {Lynch},
  \citenamefont {Back}, \citenamefont {Saini}, \citenamefont {Baisden},\ and\
  \citenamefont {McMahan}}]{Tsang1983_PLB129-18}%
  \BibitemOpen
  \bibfield  {author} {\bibinfo {author} {\bibfnamefont {M.}~\bibnamefont
  {Tsang}}, \bibinfo {author} {\bibfnamefont {H.}~\bibnamefont {Utsunomiya}},
  \bibinfo {author} {\bibfnamefont {C.}~\bibnamefont {Gelbke}}, \bibinfo
  {author} {\bibfnamefont {W.}~\bibnamefont {Lynch}}, \bibinfo {author}
  {\bibfnamefont {B.}~\bibnamefont {Back}}, \bibinfo {author} {\bibfnamefont
  {S.}~\bibnamefont {Saini}}, \bibinfo {author} {\bibfnamefont
  {P.}~\bibnamefont {Baisden}},\ and\ \bibinfo {author} {\bibfnamefont
  {M.}~\bibnamefont {McMahan}},\ }\bibfield  {title} {\bibinfo {title} {Energy
  dependence of fission fragment angular distributions for $^{19}${F},
  $^{24}${M}g and $^{28}${S}i induced reactions on $^{208}${P}b},\ }\href
  {https://doi.org/https://doi.org/10.1016/0370-2693(83)90719-0} {\bibfield
  {journal} {\bibinfo  {journal} {Phys. Lett. B}\ }\textbf {\bibinfo {volume}
  {129}},\ \bibinfo {pages} {18} (\bibinfo {year} {1983})}\BibitemShut
  {NoStop}%
\bibitem [{\citenamefont {Banerjee}\ \emph {et~al.}(2021)\citenamefont
  {Banerjee}, \citenamefont {Hinde}, \citenamefont {Dasgupta}, \citenamefont
  {Sadhukhan}, \citenamefont {Simpson}, \citenamefont {Jeung}, \citenamefont
  {Simenel}, \citenamefont {Swinton-Bland}, \citenamefont {Williams},
  \citenamefont {Bezzina}, \citenamefont {Carter}, \citenamefont {Cook},
  \citenamefont {Albers}, \citenamefont {D\"ullmann}, \citenamefont
  {Khuyagbaatar}, \citenamefont {Kindler}, \citenamefont {Lommel},
  \citenamefont {Mokry}, \citenamefont {Prasad}, \citenamefont {Runke},
  \citenamefont {Schunck}, \citenamefont {Sengupta}, \citenamefont {Smith},
  \citenamefont {Th\"orle-Pospiech}, \citenamefont {Trautmann}, \citenamefont
  {Vo-Phuoc}, \citenamefont {Walshe},\ and\ \citenamefont
  {Yakushev}}]{Banerjee2021_PLB820-136601}%
  \BibitemOpen
  \bibfield  {author} {\bibinfo {author} {\bibfnamefont {K.}~\bibnamefont
  {Banerjee}}, \bibinfo {author} {\bibfnamefont {D.}~\bibnamefont {Hinde}},
  \bibinfo {author} {\bibfnamefont {M.}~\bibnamefont {Dasgupta}}, \bibinfo
  {author} {\bibfnamefont {J.}~\bibnamefont {Sadhukhan}}, \bibinfo {author}
  {\bibfnamefont {E.}~\bibnamefont {Simpson}}, \bibinfo {author} {\bibfnamefont
  {D.}~\bibnamefont {Jeung}}, \bibinfo {author} {\bibfnamefont
  {C.}~\bibnamefont {Simenel}}, \bibinfo {author} {\bibfnamefont
  {B.}~\bibnamefont {Swinton-Bland}}, \bibinfo {author} {\bibfnamefont
  {E.}~\bibnamefont {Williams}}, \bibinfo {author} {\bibfnamefont
  {L.}~\bibnamefont {Bezzina}}, \bibinfo {author} {\bibfnamefont
  {I.}~\bibnamefont {Carter}}, \bibinfo {author} {\bibfnamefont
  {K.}~\bibnamefont {Cook}}, \bibinfo {author} {\bibfnamefont {H.}~\bibnamefont
  {Albers}}, \bibinfo {author} {\bibfnamefont {C.}~\bibnamefont {D\"ullmann}},
  \bibinfo {author} {\bibfnamefont {J.}~\bibnamefont {Khuyagbaatar}}, \bibinfo
  {author} {\bibfnamefont {B.}~\bibnamefont {Kindler}}, \bibinfo {author}
  {\bibfnamefont {B.}~\bibnamefont {Lommel}}, \bibinfo {author} {\bibfnamefont
  {C.}~\bibnamefont {Mokry}}, \bibinfo {author} {\bibfnamefont
  {E.}~\bibnamefont {Prasad}}, \bibinfo {author} {\bibfnamefont
  {J.}~\bibnamefont {Runke}}, \bibinfo {author} {\bibfnamefont
  {N.}~\bibnamefont {Schunck}}, \bibinfo {author} {\bibfnamefont
  {C.}~\bibnamefont {Sengupta}}, \bibinfo {author} {\bibfnamefont
  {J.}~\bibnamefont {Smith}}, \bibinfo {author} {\bibfnamefont
  {P.}~\bibnamefont {Th\"orle-Pospiech}}, \bibinfo {author} {\bibfnamefont
  {N.}~\bibnamefont {Trautmann}}, \bibinfo {author} {\bibfnamefont
  {K.}~\bibnamefont {Vo-Phuoc}}, \bibinfo {author} {\bibfnamefont
  {J.}~\bibnamefont {Walshe}},\ and\ \bibinfo {author} {\bibfnamefont
  {A.}~\bibnamefont {Yakushev}},\ }\bibfield  {title} {\bibinfo {title}
  {Sensitive search for near-symmetric and super-asymmetric fusion-fission of
  the superheavy element flerovium (${Z}=114$)},\ }\href
  {https://doi.org/https://doi.org/10.1016/j.physletb.2021.136601} {\bibfield
  {journal} {\bibinfo  {journal} {Phys. Lett. B}\ }\textbf {\bibinfo {volume}
  {820}},\ \bibinfo {pages} {136601} (\bibinfo {year} {2021})}\BibitemShut
  {NoStop}%
\bibitem [{\citenamefont {Hinde}\ \emph {et~al.}(2021)\citenamefont {Hinde},
  \citenamefont {Dasgupta},\ and\ \citenamefont
  {Simpson}}]{Hinde2021_PPNP118-103856}%
  \BibitemOpen
  \bibfield  {author} {\bibinfo {author} {\bibfnamefont {D.}~\bibnamefont
  {Hinde}}, \bibinfo {author} {\bibfnamefont {M.}~\bibnamefont {Dasgupta}},\
  and\ \bibinfo {author} {\bibfnamefont {E.}~\bibnamefont {Simpson}},\
  }\bibfield  {title} {\bibinfo {title} {Experimental studies of the
  competition between fusion and quasifission in the formation of heavy and
  superheavy nuclei},\ }\href {https://doi.org/10.1016/j.ppnp.2021.103856}
  {\bibfield  {journal} {\bibinfo  {journal} {Prog Part. Nucl. Phys.}\ }\textbf
  {\bibinfo {volume} {118}},\ \bibinfo {pages} {103856} (\bibinfo {year}
  {2021})}\BibitemShut {NoStop}%
\bibitem [{\citenamefont {Umar}\ and\ \citenamefont
  {Oberacker}(2006)}]{Umar2006_PRC74-021601R}%
  \BibitemOpen
  \bibfield  {author} {\bibinfo {author} {\bibfnamefont {A.~S.}\ \bibnamefont
  {Umar}}\ and\ \bibinfo {author} {\bibfnamefont {V.~E.}\ \bibnamefont
  {Oberacker}},\ }\bibfield  {title} {\bibinfo {title} {Heavy-ion interaction
  potential deduced from density-constrained time-dependent {H}artree-{F}ock
  calculation},\ }\href {https://doi.org/10.1103/PhysRevC.74.021601} {\bibfield
   {journal} {\bibinfo  {journal} {Phys. Rev. C}\ }\textbf {\bibinfo {volume}
  {74}},\ \bibinfo {pages} {021601(R)} (\bibinfo {year} {2006})}\BibitemShut
  {NoStop}%
\bibitem [{\citenamefont {Wen}\ \emph {et~al.}(2013)\citenamefont {Wen},
  \citenamefont {Sakata}, \citenamefont {Li}, \citenamefont {Wu}, \citenamefont
  {Zhang},\ and\ \citenamefont {Zhou}}]{Wen2013_PRL111-12501}%
  \BibitemOpen
  \bibfield  {author} {\bibinfo {author} {\bibfnamefont {K.}~\bibnamefont
  {Wen}}, \bibinfo {author} {\bibfnamefont {F.}~\bibnamefont {Sakata}},
  \bibinfo {author} {\bibfnamefont {Z.-X.}\ \bibnamefont {Li}}, \bibinfo
  {author} {\bibfnamefont {X.-Z.}\ \bibnamefont {Wu}}, \bibinfo {author}
  {\bibfnamefont {Y.-X.}\ \bibnamefont {Zhang}},\ and\ \bibinfo {author}
  {\bibfnamefont {S.-G.}\ \bibnamefont {Zhou}},\ }\bibfield  {title} {\bibinfo
  {title} {Non-gaussian fluctuations and non-markovian effects in the nuclear
  fusion process: Langevin dynamics emerging from quantum molecular dynamics
  simulations},\ }\href {https://doi.org/10.1103/PhysRevLett.111.012501}
  {\bibfield  {journal} {\bibinfo  {journal} {Phys. Rev. Lett.}\ }\textbf
  {\bibinfo {volume} {111}},\ \bibinfo {pages} {012501} (\bibinfo {year}
  {2013})}\BibitemShut {NoStop}%
\bibitem [{\citenamefont {Simenel}\ \emph
  {et~al.}(2013{\natexlab{a}})\citenamefont {Simenel}, \citenamefont
  {Dasgupta}, \citenamefont {Hinde},\ and\ \citenamefont
  {Williams}}]{Simenel2013_PRC88-064604}%
  \BibitemOpen
  \bibfield  {author} {\bibinfo {author} {\bibfnamefont {C.}~\bibnamefont
  {Simenel}}, \bibinfo {author} {\bibfnamefont {M.}~\bibnamefont {Dasgupta}},
  \bibinfo {author} {\bibfnamefont {D.~J.}\ \bibnamefont {Hinde}},\ and\
  \bibinfo {author} {\bibfnamefont {E.}~\bibnamefont {Williams}},\ }\bibfield
  {title} {\bibinfo {title} {Microscopic approach to coupled-channels effects
  on fusion},\ }\href {https://doi.org/10.1103/PhysRevC.88.064604} {\bibfield
  {journal} {\bibinfo  {journal} {Phys. Rev. C}\ }\textbf {\bibinfo {volume}
  {88}},\ \bibinfo {pages} {064604} (\bibinfo {year}
  {2013}{\natexlab{a}})}\BibitemShut {NoStop}%
\bibitem [{\citenamefont {Washiyama}\ and\ \citenamefont
  {Lacroix}(2008)}]{Washiyama2008_PRC78-024610}%
  \BibitemOpen
  \bibfield  {author} {\bibinfo {author} {\bibfnamefont {K.}~\bibnamefont
  {Washiyama}}\ and\ \bibinfo {author} {\bibfnamefont {D.}~\bibnamefont
  {Lacroix}},\ }\bibfield  {title} {\bibinfo {title} {Energy dependence of the
  nucleus-nucleus potential close to the {C}oulomb barrier},\ }\href
  {https://doi.org/10.1103/PhysRevC.78.024610} {\bibfield  {journal} {\bibinfo
  {journal} {Phys. Rev. C}\ }\textbf {\bibinfo {volume} {78}},\ \bibinfo
  {pages} {024610} (\bibinfo {year} {2008})}\BibitemShut {NoStop}%
\bibitem [{\citenamefont {Simenel}(2012)}]{Simenel2012_EPJA48-152}%
  \BibitemOpen
  \bibfield  {author} {\bibinfo {author} {\bibfnamefont {C.}~\bibnamefont
  {Simenel}},\ }\bibfield  {title} {\bibinfo {title} {Nuclear quantum many-body
  dynamics},\ }\href {https://doi.org/10.1140/epja/i2012-12152-0} {\bibfield
  {journal} {\bibinfo  {journal} {Eur. Phys. J. A}\ }\textbf {\bibinfo {volume}
  {48}},\ \bibinfo {pages} {152} (\bibinfo {year} {2012})}\BibitemShut
  {NoStop}%
\bibitem [{\citenamefont {Nakatsukasa}\ \emph {et~al.}(2016)\citenamefont
  {Nakatsukasa}, \citenamefont {Matsuyanagi}, \citenamefont {Matsuo},\ and\
  \citenamefont {Yabana}}]{Nakatsukasa2016_RMP88-45004}%
  \BibitemOpen
  \bibfield  {author} {\bibinfo {author} {\bibfnamefont {T.}~\bibnamefont
  {Nakatsukasa}}, \bibinfo {author} {\bibfnamefont {K.}~\bibnamefont
  {Matsuyanagi}}, \bibinfo {author} {\bibfnamefont {M.}~\bibnamefont
  {Matsuo}},\ and\ \bibinfo {author} {\bibfnamefont {K.}~\bibnamefont
  {Yabana}},\ }\bibfield  {title} {\bibinfo {title} {Time-dependent
  density-functional description of nuclear dynamics},\ }\href
  {https://doi.org/10.1103/RevModPhys.88.045004} {\bibfield  {journal}
  {\bibinfo  {journal} {Rev. Mod. Phys.}\ }\textbf {\bibinfo {volume} {88}},\
  \bibinfo {pages} {045004} (\bibinfo {year} {2016})}\BibitemShut {NoStop}%
\bibitem [{\citenamefont {Simenel}\ and\ \citenamefont
  {Umar}(2018)}]{Simenel2018_PPNP103-19}%
  \BibitemOpen
  \bibfield  {author} {\bibinfo {author} {\bibfnamefont {C.}~\bibnamefont
  {Simenel}}\ and\ \bibinfo {author} {\bibfnamefont {A.}~\bibnamefont {Umar}},\
  }\bibfield  {title} {\bibinfo {title} {Heavy--ion collisions and fission
  dynamics with the time--dependent {H}artree--{F}ock theory and its
  extensions},\ }\href {https://doi.org/10.1016/j.ppnp.2018.07.002} {\bibfield
  {journal} {\bibinfo  {journal} {Prog. Part. Nucl. Phys.}\ }\textbf {\bibinfo
  {volume} {103}},\ \bibinfo {pages} {19} (\bibinfo {year} {2018})}\BibitemShut
  {NoStop}%
\bibitem [{\citenamefont {Stevenson}\ and\ \citenamefont
  {Barton}(2019)}]{Stevenson2019_PPNP104-142164}%
  \BibitemOpen
  \bibfield  {author} {\bibinfo {author} {\bibfnamefont {P.}~\bibnamefont
  {Stevenson}}\ and\ \bibinfo {author} {\bibfnamefont {M.}~\bibnamefont
  {Barton}},\ }\bibfield  {title} {\bibinfo {title} {Low-energy heavy-ion
  reactions and the {S}kyrme effective interaction},\ }\href
  {https://doi.org/https://doi.org/10.1016/j.ppnp.2018.09.002} {\bibfield
  {journal} {\bibinfo  {journal} {Prog. Part. Nucl. Phys.}\ }\textbf {\bibinfo
  {volume} {104}},\ \bibinfo {pages} {142} (\bibinfo {year}
  {2019})}\BibitemShut {NoStop}%
\bibitem [{\citenamefont {Sekizawa}(2019)}]{Sekizawa2019_FP7-20}%
  \BibitemOpen
  \bibfield  {author} {\bibinfo {author} {\bibfnamefont {K.}~\bibnamefont
  {Sekizawa}},\ }\bibfield  {title} {\bibinfo {title} {{T}{D}{H}{F} theory and
  its extensions for the multinucleon transfer reaction: A mini review},\
  }\href {https://doi.org/10.3389/fphy.2019.00020} {\bibfield  {journal}
  {\bibinfo  {journal} {Front. Phys.}\ }\textbf {\bibinfo {volume} {7}},\
  \bibinfo {pages} {20} (\bibinfo {year} {2019})}\BibitemShut {NoStop}%
\bibitem [{\citenamefont {Guo}\ \emph {et~al.}(2018{\natexlab{a}})\citenamefont
  {Guo}, \citenamefont {Shen}, \citenamefont {Yu},\ and\ \citenamefont
  {Wu}}]{Guo2018_PRC98-064609}%
  \BibitemOpen
  \bibfield  {author} {\bibinfo {author} {\bibfnamefont {L.}~\bibnamefont
  {Guo}}, \bibinfo {author} {\bibfnamefont {C.}~\bibnamefont {Shen}}, \bibinfo
  {author} {\bibfnamefont {C.}~\bibnamefont {Yu}},\ and\ \bibinfo {author}
  {\bibfnamefont {Z.}~\bibnamefont {Wu}},\ }\bibfield  {title} {\bibinfo
  {title} {Isotopic trends of quasifission and fusion-fission in the reactions
  $^{48}\mathrm{Ca}+^{239,244}\mathrm{Pu}$},\ }\href
  {https://doi.org/10.1103/PhysRevC.98.064609} {\bibfield  {journal} {\bibinfo
  {journal} {Phys. Rev. C}\ }\textbf {\bibinfo {volume} {98}},\ \bibinfo
  {pages} {064609} (\bibinfo {year} {2018}{\natexlab{a}})}\BibitemShut
  {NoStop}%
\bibitem [{\citenamefont {Simenel}\ \emph
  {et~al.}(2013{\natexlab{b}})\citenamefont {Simenel}, \citenamefont {Keser},
  \citenamefont {Umar},\ and\ \citenamefont
  {Oberacker}}]{Simenel2013_PRC88-024617}%
  \BibitemOpen
  \bibfield  {author} {\bibinfo {author} {\bibfnamefont {C.}~\bibnamefont
  {Simenel}}, \bibinfo {author} {\bibfnamefont {R.}~\bibnamefont {Keser}},
  \bibinfo {author} {\bibfnamefont {A.~S.}\ \bibnamefont {Umar}},\ and\
  \bibinfo {author} {\bibfnamefont {V.~E.}\ \bibnamefont {Oberacker}},\
  }\bibfield  {title} {\bibinfo {title} {Microscopic study of
  ${}^{16}\mathrm{O}+{}^{16}\mathrm{O}$ fusion},\ }\href
  {https://doi.org/10.1103/PhysRevC.88.024617} {\bibfield  {journal} {\bibinfo
  {journal} {Phys. Rev. C}\ }\textbf {\bibinfo {volume} {88}},\ \bibinfo
  {pages} {024617} (\bibinfo {year} {2013}{\natexlab{b}})}\BibitemShut
  {NoStop}%
\bibitem [{\citenamefont {Guo}\ \emph {et~al.}(2018{\natexlab{b}})\citenamefont
  {Guo}, \citenamefont {Simenel}, \citenamefont {Shi},\ and\ \citenamefont
  {Yu}}]{Guo2018_PLB782-401}%
  \BibitemOpen
  \bibfield  {author} {\bibinfo {author} {\bibfnamefont {L.}~\bibnamefont
  {Guo}}, \bibinfo {author} {\bibfnamefont {C.}~\bibnamefont {Simenel}},
  \bibinfo {author} {\bibfnamefont {L.}~\bibnamefont {Shi}},\ and\ \bibinfo
  {author} {\bibfnamefont {C.}~\bibnamefont {Yu}},\ }\bibfield  {title}
  {\bibinfo {title} {The role of tensor force in heavy-ion fusion dynamics},\
  }\href {https://doi.org/https://doi.org/10.1016/j.physletb.2018.05.066}
  {\bibfield  {journal} {\bibinfo  {journal} {Phys. Lett. B}\ }\textbf
  {\bibinfo {volume} {782}},\ \bibinfo {pages} {401 } (\bibinfo {year}
  {2018}{\natexlab{b}})}\BibitemShut {NoStop}%
\bibitem [{\citenamefont {Sekizawa}\ and\ \citenamefont
  {Hagino}(2019)}]{Sekizawa2019_PRC99-051602R}%
  \BibitemOpen
  \bibfield  {author} {\bibinfo {author} {\bibfnamefont {K.}~\bibnamefont
  {Sekizawa}}\ and\ \bibinfo {author} {\bibfnamefont {K.}~\bibnamefont
  {Hagino}},\ }\bibfield  {title} {\bibinfo {title} {Time--dependent
  {H}artree--{F}ock plus {L}angevin approach for hot fusion reactions to
  synthesize the ${Z}=120$ superheavy element},\ }\href
  {https://doi.org/10.1103/PhysRevC.99.051602} {\bibfield  {journal} {\bibinfo
  {journal} {Phys. Rev. C}\ }\textbf {\bibinfo {volume} {99}},\ \bibinfo
  {pages} {051602(R)} (\bibinfo {year} {2019})}\BibitemShut {NoStop}%
\bibitem [{\citenamefont {Guo}\ and\ \citenamefont
  {Nakatsukasa}(2012)}]{Guo2012_EWC38-09003}%
  \BibitemOpen
  \bibfield  {author} {\bibinfo {author} {\bibfnamefont {L.}~\bibnamefont
  {Guo}}\ and\ \bibinfo {author} {\bibfnamefont {T.}~\bibnamefont
  {Nakatsukasa}},\ }\bibfield  {title} {\bibinfo {title} {Time-dependent
  {H}artree-{F}ock studies of the dynamical fusion threshold},\ }\href
  {https://doi.org/10.1051/epjconf/20123809003} {\bibfield  {journal} {\bibinfo
   {journal} {{EPJ} Web of Conferences}\ }\textbf {\bibinfo {volume} {38}},\
  \bibinfo {pages} {09003} (\bibinfo {year} {2012})}\BibitemShut {NoStop}%
\bibitem [{\citenamefont {Umar}\ and\ \citenamefont
  {Oberacker}(2008)}]{Umar2008_PRC77-064605}%
  \BibitemOpen
  \bibfield  {author} {\bibinfo {author} {\bibfnamefont {A.~S.}\ \bibnamefont
  {Umar}}\ and\ \bibinfo {author} {\bibfnamefont {V.~E.}\ \bibnamefont
  {Oberacker}},\ }\bibfield  {title} {\bibinfo {title}
  {${}^{64}\mathrm{Ni}{+}^{64}\mathrm{Ni}$ fusion reaction calculated with the
  density-constrained time-dependent {H}artree-{F}ock formalism},\ }\href
  {https://doi.org/10.1103/PhysRevC.77.064605} {\bibfield  {journal} {\bibinfo
  {journal} {Phys. Rev. C}\ }\textbf {\bibinfo {volume} {77}},\ \bibinfo
  {pages} {064605} (\bibinfo {year} {2008})}\BibitemShut {NoStop}%
\bibitem [{\citenamefont {Umar}\ \emph {et~al.}(2010)\citenamefont {Umar},
  \citenamefont {Oberacker}, \citenamefont {Maruhn},\ and\ \citenamefont
  {Reinhard}}]{Umar2010_PRC81-064607}%
  \BibitemOpen
  \bibfield  {author} {\bibinfo {author} {\bibfnamefont {A.~S.}\ \bibnamefont
  {Umar}}, \bibinfo {author} {\bibfnamefont {V.~E.}\ \bibnamefont {Oberacker}},
  \bibinfo {author} {\bibfnamefont {J.~A.}\ \bibnamefont {Maruhn}},\ and\
  \bibinfo {author} {\bibfnamefont {P.-G.}\ \bibnamefont {Reinhard}},\
  }\bibfield  {title} {\bibinfo {title} {Entrance channel dynamics of hot and
  cold fusion reactions leading to superheavy elements},\ }\href
  {https://doi.org/10.1103/PhysRevC.81.064607} {\bibfield  {journal} {\bibinfo
  {journal} {Phys. Rev. C}\ }\textbf {\bibinfo {volume} {81}},\ \bibinfo
  {pages} {064607} (\bibinfo {year} {2010})}\BibitemShut {NoStop}%
\bibitem [{\citenamefont {Umar}\ \emph {et~al.}(2014)\citenamefont {Umar},
  \citenamefont {Simenel},\ and\ \citenamefont
  {Oberacker}}]{Umar2014_PRC89-034611}%
  \BibitemOpen
  \bibfield  {author} {\bibinfo {author} {\bibfnamefont {A.}~\bibnamefont
  {Umar}}, \bibinfo {author} {\bibfnamefont {C.}~\bibnamefont {Simenel}},\ and\
  \bibinfo {author} {\bibfnamefont {V.}~\bibnamefont {Oberacker}},\ }\bibfield
  {title} {\bibinfo {title} {Energy dependence of potential barriers and its
  effect on fusion cross-sections},\ }\href
  {https://doi.org/10.1103/PhysRevC.89.034611} {\bibfield  {journal} {\bibinfo
  {journal} {Phys. Rev. C}\ }\textbf {\bibinfo {volume} {89}},\ \bibinfo
  {pages} {034611} (\bibinfo {year} {2014})}\BibitemShut {NoStop}%
\bibitem [{\citenamefont {Umar}\ \emph {et~al.}(2016)\citenamefont {Umar},
  \citenamefont {Oberacker},\ and\ \citenamefont
  {Simenel}}]{Umar2016_PRC94-024605}%
  \BibitemOpen
  \bibfield  {author} {\bibinfo {author} {\bibfnamefont {A.}~\bibnamefont
  {Umar}}, \bibinfo {author} {\bibfnamefont {V.}~\bibnamefont {Oberacker}},\
  and\ \bibinfo {author} {\bibfnamefont {C.}~\bibnamefont {Simenel}},\
  }\bibfield  {title} {\bibinfo {title} {Fusion and quasifission dynamics in
  the reactions $^{48}${C}a+$^{249}${B}k and $^{50}${T}i+$^{249}${B}k using
  {T}{D}{H}{F}},\ }\href {https://doi.org/10.1103/PhysRevC.94.024605}
  {\bibfield  {journal} {\bibinfo  {journal} {Phys. Rev. C}\ }\textbf {\bibinfo
  {volume} {94}},\ \bibinfo {pages} {024605} (\bibinfo {year}
  {2016})}\BibitemShut {NoStop}%
\bibitem [{\citenamefont {Guo}\ \emph {et~al.}(2018{\natexlab{c}})\citenamefont
  {Guo}, \citenamefont {Godbey},\ and\ \citenamefont
  {Umar}}]{Guo2018_PRC98-064607}%
  \BibitemOpen
  \bibfield  {author} {\bibinfo {author} {\bibfnamefont {L.}~\bibnamefont
  {Guo}}, \bibinfo {author} {\bibfnamefont {K.}~\bibnamefont {Godbey}},\ and\
  \bibinfo {author} {\bibfnamefont {A.~S.}\ \bibnamefont {Umar}},\ }\bibfield
  {title} {\bibinfo {title} {Influence of the tensor force on the microscopic
  heavy-ion interaction potential},\ }\href
  {https://doi.org/10.1103/PhysRevC.98.064607} {\bibfield  {journal} {\bibinfo
  {journal} {Phys. Rev. C}\ }\textbf {\bibinfo {volume} {98}},\ \bibinfo
  {pages} {064607} (\bibinfo {year} {2018}{\natexlab{c}})}\BibitemShut
  {NoStop}%
\bibitem [{\citenamefont {Godbey}\ \emph {et~al.}(2019)\citenamefont {Godbey},
  \citenamefont {Guo},\ and\ \citenamefont {Umar}}]{Godbey2019_PRC100-054612}%
  \BibitemOpen
  \bibfield  {author} {\bibinfo {author} {\bibfnamefont {K.}~\bibnamefont
  {Godbey}}, \bibinfo {author} {\bibfnamefont {L.}~\bibnamefont {Guo}},\ and\
  \bibinfo {author} {\bibfnamefont {A.~S.}\ \bibnamefont {Umar}},\ }\bibfield
  {title} {\bibinfo {title} {Influence of the tensor interaction on heavy-ion
  fusion cross sections},\ }\href {https://doi.org/10.1103/PhysRevC.100.054612}
  {\bibfield  {journal} {\bibinfo  {journal} {Phys. Rev. C}\ }\textbf {\bibinfo
  {volume} {100}},\ \bibinfo {pages} {054612} (\bibinfo {year}
  {2019})}\BibitemShut {NoStop}%
\bibitem [{\citenamefont {Simenel}\ \emph {et~al.}(2017)\citenamefont
  {Simenel}, \citenamefont {Umar}, \citenamefont {Godbey}, \citenamefont
  {Dasgupta},\ and\ \citenamefont {Hinde}}]{Simenel2017_PRC95-031601R}%
  \BibitemOpen
  \bibfield  {author} {\bibinfo {author} {\bibfnamefont {C.}~\bibnamefont
  {Simenel}}, \bibinfo {author} {\bibfnamefont {A.~S.}\ \bibnamefont {Umar}},
  \bibinfo {author} {\bibfnamefont {K.}~\bibnamefont {Godbey}}, \bibinfo
  {author} {\bibfnamefont {M.}~\bibnamefont {Dasgupta}},\ and\ \bibinfo
  {author} {\bibfnamefont {D.~J.}\ \bibnamefont {Hinde}},\ }\bibfield  {title}
  {\bibinfo {title} {How the {P}auli exclusion principle affects fusion of
  atomic nuclei},\ }\href {https://doi.org/10.1103/PhysRevC.95.031601}
  {\bibfield  {journal} {\bibinfo  {journal} {Phys. Rev. C}\ }\textbf {\bibinfo
  {volume} {95}},\ \bibinfo {pages} {031601(R)} (\bibinfo {year}
  {2017})}\BibitemShut {NoStop}%
\bibitem [{\citenamefont {Wang}\ \emph {et~al.}(2006)\citenamefont {Wang},
  \citenamefont {Wu}, \citenamefont {Li}, \citenamefont {Liu},\ and\
  \citenamefont {Scheid}}]{Wang2006_PRC74-044604}%
  \BibitemOpen
  \bibfield  {author} {\bibinfo {author} {\bibfnamefont {N.}~\bibnamefont
  {Wang}}, \bibinfo {author} {\bibfnamefont {X.}~\bibnamefont {Wu}}, \bibinfo
  {author} {\bibfnamefont {Z.}~\bibnamefont {Li}}, \bibinfo {author}
  {\bibfnamefont {M.}~\bibnamefont {Liu}},\ and\ \bibinfo {author}
  {\bibfnamefont {W.}~\bibnamefont {Scheid}},\ }\bibfield  {title} {\bibinfo
  {title} {Applications of {S}kyrme energy-density functional to fusion
  reactions for synthesis of superheavy nuclei},\ }\href
  {https://doi.org/10.1103/PhysRevC.74.044604} {\bibfield  {journal} {\bibinfo
  {journal} {Phys. Rev. C}\ }\textbf {\bibinfo {volume} {74}},\ \bibinfo
  {pages} {044604} (\bibinfo {year} {2006})}\BibitemShut {NoStop}%
\bibitem [{\citenamefont {Hagino}\ \emph {et~al.}(1999)\citenamefont {Hagino},
  \citenamefont {Rowley},\ and\ \citenamefont
  {Kruppa}}]{Hagino1999_CPC123-143}%
  \BibitemOpen
  \bibfield  {author} {\bibinfo {author} {\bibfnamefont {K.}~\bibnamefont
  {Hagino}}, \bibinfo {author} {\bibfnamefont {N.}~\bibnamefont {Rowley}},\
  and\ \bibinfo {author} {\bibfnamefont {A.~T.}\ \bibnamefont {Kruppa}},\
  }\bibfield  {title} {\bibinfo {title} {{A program for coupled-channel
  calculations with all order couplings for heavy-ion fusion reactions}},\
  }\href {https://doi.org/10.1016/S0010-4655(99)00243-X} {\bibfield  {journal}
  {\bibinfo  {journal} {Comput. Phys. Commun.}\ }\textbf {\bibinfo {volume}
  {123}},\ \bibinfo {pages} {143} (\bibinfo {year} {1999})}\BibitemShut
  {NoStop}%
\bibitem [{\citenamefont {Wong}(1973)}]{Wong1973_PRL31-766}%
  \BibitemOpen
  \bibfield  {author} {\bibinfo {author} {\bibfnamefont {C.~Y.}\ \bibnamefont
  {Wong}},\ }\bibfield  {title} {\bibinfo {title} {Interaction barrier in
  charged-particle nuclear reactions},\ }\href
  {https://doi.org/10.1103/PhysRevLett.31.766} {\bibfield  {journal} {\bibinfo
  {journal} {Phys. Rev. Lett.}\ }\textbf {\bibinfo {volume} {31}},\ \bibinfo
  {pages} {766} (\bibinfo {year} {1973})}\BibitemShut {NoStop}%
\bibitem [{\citenamefont {Oberacker}\ \emph {et~al.}(2014)\citenamefont
  {Oberacker}, \citenamefont {Umar},\ and\ \citenamefont
  {Simenel}}]{Oberacker2014_PRC90-054605}%
  \BibitemOpen
  \bibfield  {author} {\bibinfo {author} {\bibfnamefont {V.~E.}\ \bibnamefont
  {Oberacker}}, \bibinfo {author} {\bibfnamefont {A.~S.}\ \bibnamefont
  {Umar}},\ and\ \bibinfo {author} {\bibfnamefont {C.}~\bibnamefont
  {Simenel}},\ }\bibfield  {title} {\bibinfo {title} {Dissipative dynamics in
  quasifission},\ }\href {https://doi.org/10.1103/PhysRevC.90.054605}
  {\bibfield  {journal} {\bibinfo  {journal} {Phys. Rev. C}\ }\textbf {\bibinfo
  {volume} {90}},\ \bibinfo {pages} {054605} (\bibinfo {year}
  {2014})}\BibitemShut {NoStop}%
\bibitem [{\citenamefont {Wakhle}\ \emph {et~al.}(2014)\citenamefont {Wakhle},
  \citenamefont {Simenel}, \citenamefont {Hinde}, \citenamefont {Dasgupta},
  \citenamefont {Evers}, \citenamefont {Luong}, \citenamefont {du~Rietz},\ and\
  \citenamefont {Williams}}]{Wakhle2014_PRL113-182502}%
  \BibitemOpen
  \bibfield  {author} {\bibinfo {author} {\bibfnamefont {A.}~\bibnamefont
  {Wakhle}}, \bibinfo {author} {\bibfnamefont {C.}~\bibnamefont {Simenel}},
  \bibinfo {author} {\bibfnamefont {D.~J.}\ \bibnamefont {Hinde}}, \bibinfo
  {author} {\bibfnamefont {M.}~\bibnamefont {Dasgupta}}, \bibinfo {author}
  {\bibfnamefont {M.}~\bibnamefont {Evers}}, \bibinfo {author} {\bibfnamefont
  {D.~H.}\ \bibnamefont {Luong}}, \bibinfo {author} {\bibfnamefont
  {R.}~\bibnamefont {du~Rietz}},\ and\ \bibinfo {author} {\bibfnamefont
  {E.}~\bibnamefont {Williams}},\ }\bibfield  {title} {\bibinfo {title}
  {Interplay between quantum shells and orientation in quasifission},\ }\href
  {https://doi.org/10.1103/PhysRevLett.113.182502} {\bibfield  {journal}
  {\bibinfo  {journal} {Phys. Rev. Lett.}\ }\textbf {\bibinfo {volume} {113}},\
  \bibinfo {pages} {182502} (\bibinfo {year} {2014})}\BibitemShut {NoStop}%
\bibitem [{\citenamefont {Umar}\ \emph {et~al.}(2015)\citenamefont {Umar},
  \citenamefont {Oberacker},\ and\ \citenamefont
  {Simenel}}]{Umar2015_PRC92-024621}%
  \BibitemOpen
  \bibfield  {author} {\bibinfo {author} {\bibfnamefont {A.~S.}\ \bibnamefont
  {Umar}}, \bibinfo {author} {\bibfnamefont {V.~E.}\ \bibnamefont
  {Oberacker}},\ and\ \bibinfo {author} {\bibfnamefont {C.}~\bibnamefont
  {Simenel}},\ }\bibfield  {title} {\bibinfo {title} {Shape evolution and
  collective dynamics of quasifission in the time-dependent {H}artree-{F}ock
  approach},\ }\href {https://doi.org/10.1103/PhysRevC.92.024621} {\bibfield
  {journal} {\bibinfo  {journal} {Phys. Rev. C}\ }\textbf {\bibinfo {volume}
  {92}},\ \bibinfo {pages} {024621} (\bibinfo {year} {2015})}\BibitemShut
  {NoStop}%
\bibitem [{\citenamefont {Sekizawa}\ and\ \citenamefont
  {Yabana}(2016)}]{Sekizawa2016_PRC93-054616}%
  \BibitemOpen
  \bibfield  {author} {\bibinfo {author} {\bibfnamefont {K.}~\bibnamefont
  {Sekizawa}}\ and\ \bibinfo {author} {\bibfnamefont {K.}~\bibnamefont
  {Yabana}},\ }\bibfield  {title} {\bibinfo {title} {Time-dependent
  {H}artree-{F}ock calculations for multinucleon transfer and quasifission
  processes in the $^{64}\text{Ni}+^{238}\text{U}$ reaction},\ }\href
  {https://doi.org/10.1103/PhysRevC.93.054616} {\bibfield  {journal} {\bibinfo
  {journal} {Phys. Rev. C}\ }\textbf {\bibinfo {volume} {93}},\ \bibinfo
  {pages} {054616} (\bibinfo {year} {2016})}\BibitemShut {NoStop}%
\bibitem [{\citenamefont {\'{S}wi\k{a}tecki}\ \emph {et~al.}(2003)\citenamefont
  {\'{S}wi\k{a}tecki}, \citenamefont {Siwek-Wilczy\'{n}ska},\ and\
  \citenamefont {Wilczy\'{n}ski}}]{Swiatecki2003_APPB34-2049}%
  \BibitemOpen
  \bibfield  {author} {\bibinfo {author} {\bibfnamefont {W.~J.}\ \bibnamefont
  {\'{S}wi\k{a}tecki}}, \bibinfo {author} {\bibfnamefont {K.}~\bibnamefont
  {Siwek-Wilczy\'{n}ska}},\ and\ \bibinfo {author} {\bibfnamefont
  {J.}~\bibnamefont {Wilczy\'{n}ski}},\ }\bibfield  {title} {\bibinfo {title}
  {Fusion by diffusion},\ }\href {https://www.actaphys.uj.edu.pl/R/34/4/2049}
  {\bibfield  {journal} {\bibinfo  {journal} {Acta Phys. Pol. B}\ }\textbf
  {\bibinfo {volume} {34}},\ \bibinfo {pages} {2049} (\bibinfo {year}
  {2003})}\BibitemShut {NoStop}%
\bibitem [{\citenamefont {Siwek-Wilczy\'{n}ska}\ \emph
  {et~al.}(2012)\citenamefont {Siwek-Wilczy\'{n}ska}, \citenamefont {Cap},
  \citenamefont {Kowal}, \citenamefont {Sobiczewski},\ and\ \citenamefont
  {Wilczy\'{n}ski}}]{SiwekWilczynska2012_PRC86-014611}%
  \BibitemOpen
  \bibfield  {author} {\bibinfo {author} {\bibfnamefont {K.}~\bibnamefont
  {Siwek-Wilczy\'{n}ska}}, \bibinfo {author} {\bibfnamefont {T.}~\bibnamefont
  {Cap}}, \bibinfo {author} {\bibfnamefont {M.}~\bibnamefont {Kowal}}, \bibinfo
  {author} {\bibfnamefont {A.}~\bibnamefont {Sobiczewski}},\ and\ \bibinfo
  {author} {\bibfnamefont {J.}~\bibnamefont {Wilczy\'{n}ski}},\ }\bibfield
  {title} {\bibinfo {title} {Predictions of the fusion-by-diffusion model for
  the synthesis cross sections of ${Z}=114$--120 elements based on
  macroscopic-microscopic fission barriers},\ }\href
  {https://doi.org/10.1103/PhysRevC.86.014611} {\bibfield  {journal} {\bibinfo
  {journal} {Phys. Rev. C}\ }\textbf {\bibinfo {volume} {86}},\ \bibinfo
  {pages} {014611} (\bibinfo {year} {2012})}\BibitemShut {NoStop}%
\bibitem [{\citenamefont {Maruhn}\ \emph {et~al.}(2014)\citenamefont {Maruhn},
  \citenamefont {Reinhard}, \citenamefont {Stevenson},\ and\ \citenamefont
  {Umar}}]{Maruhn2014_CPC185-2195}%
  \BibitemOpen
  \bibfield  {author} {\bibinfo {author} {\bibfnamefont {J.}~\bibnamefont
  {Maruhn}}, \bibinfo {author} {\bibfnamefont {P.-G.}\ \bibnamefont
  {Reinhard}}, \bibinfo {author} {\bibfnamefont {P.}~\bibnamefont
  {Stevenson}},\ and\ \bibinfo {author} {\bibfnamefont {A.}~\bibnamefont
  {Umar}},\ }\bibfield  {title} {\bibinfo {title} {The tdhf code sky3d},\
  }\href {https://doi.org/https://doi.org/10.1016/j.cpc.2014.04.008} {\bibfield
   {journal} {\bibinfo  {journal} {Comp. Phys. Commun.}\ }\textbf {\bibinfo
  {volume} {185}},\ \bibinfo {pages} {2195} (\bibinfo {year}
  {2014})}\BibitemShut {NoStop}%
\bibitem [{\citenamefont {Dai}\ \emph {et~al.}(2014)\citenamefont {Dai},
  \citenamefont {Guo}, \citenamefont {Zhao},\ and\ \citenamefont
  {Zhou}}]{Dai2014_PRC90-044609}%
  \BibitemOpen
  \bibfield  {author} {\bibinfo {author} {\bibfnamefont {G.-F.}\ \bibnamefont
  {Dai}}, \bibinfo {author} {\bibfnamefont {L.}~\bibnamefont {Guo}}, \bibinfo
  {author} {\bibfnamefont {E.-G.}\ \bibnamefont {Zhao}},\ and\ \bibinfo
  {author} {\bibfnamefont {S.-G.}\ \bibnamefont {Zhou}},\ }\bibfield  {title}
  {\bibinfo {title} {Dissipation dynamics and spin-orbit force in
  time-dependent {H}artree-{F}ock theory},\ }\href
  {https://doi.org/10.1103/PhysRevC.90.044609} {\bibfield  {journal} {\bibinfo
  {journal} {Phys. Rev. C}\ }\textbf {\bibinfo {volume} {90}},\ \bibinfo
  {pages} {044609} (\bibinfo {year} {2014})}\BibitemShut {NoStop}%
\bibitem [{\citenamefont {Wu}\ and\ \citenamefont
  {Guo}(2019)}]{Wu2019_PRC100-014612}%
  \BibitemOpen
  \bibfield  {author} {\bibinfo {author} {\bibfnamefont {Z.}~\bibnamefont
  {Wu}}\ and\ \bibinfo {author} {\bibfnamefont {L.}~\bibnamefont {Guo}},\
  }\bibfield  {title} {\bibinfo {title} {Microscopic studies of production
  cross sections in multinucleon transfer reaction
  $^{58}\mathrm{Ni}+^{124}\mathrm{Sn}$},\ }\href
  {https://doi.org/10.1103/PhysRevC.100.014612} {\bibfield  {journal} {\bibinfo
   {journal} {Phys. Rev. C}\ }\textbf {\bibinfo {volume} {100}},\ \bibinfo
  {pages} {014612} (\bibinfo {year} {2019})}\BibitemShut {NoStop}%
\bibitem [{\citenamefont {Li}\ \emph {et~al.}(2019)\citenamefont {Li},
  \citenamefont {Wu},\ and\ \citenamefont {Guo}}]{Li2019_SCPMA62-122011}%
  \BibitemOpen
  \bibfield  {author} {\bibinfo {author} {\bibfnamefont {X.}~\bibnamefont
  {Li}}, \bibinfo {author} {\bibfnamefont {Z.}~\bibnamefont {Wu}},\ and\
  \bibinfo {author} {\bibfnamefont {L.}~\bibnamefont {Guo}},\ }\bibfield
  {title} {\bibinfo {title} {Entrance-channel dynamics in the reaction
  $^{40}${C}a+$^{208}${P}b},\ }\href
  {https://doi.org/10.1007/s11433-019-9435-x} {\bibfield  {journal} {\bibinfo
  {journal} {Sci. China: Phys. Mech. Astron}\ }\textbf {\bibinfo {volume}
  {62}},\ \bibinfo {pages} {122011} (\bibinfo {year} {2019})}\BibitemShut
  {NoStop}%
\bibitem [{\citenamefont {Wu}\ and\ \citenamefont
  {Guo}(2020)}]{Wu2020_SCPMA63-242021}%
  \BibitemOpen
  \bibfield  {author} {\bibinfo {author} {\bibfnamefont {Z.}~\bibnamefont
  {Wu}}\ and\ \bibinfo {author} {\bibfnamefont {L.}~\bibnamefont {Guo}},\
  }\bibfield  {title} {\bibinfo {title} {Production of proton-rich actinide
  nuclei in the multinucleon transfer reaction $^{58}${N}i+$^{232}${T}h},\
  }\href {https://doi.org/10.1007/s11433-019-1484-0} {\bibfield  {journal}
  {\bibinfo  {journal} {Sci. China Phys. Mech. Astron.}\ }\textbf {\bibinfo
  {volume} {63}},\ \bibinfo {pages} {242021} (\bibinfo {year}
  {2020})}\BibitemShut {NoStop}%
\bibitem [{\citenamefont {Wu}\ \emph {et~al.}(2022)\citenamefont {Wu},
  \citenamefont {Guo}, \citenamefont {Liu},\ and\ \citenamefont
  {Peng}}]{Wu2022_PLB825-136886}%
  \BibitemOpen
  \bibfield  {author} {\bibinfo {author} {\bibfnamefont {Z.}~\bibnamefont
  {Wu}}, \bibinfo {author} {\bibfnamefont {L.}~\bibnamefont {Guo}}, \bibinfo
  {author} {\bibfnamefont {Z.}~\bibnamefont {Liu}},\ and\ \bibinfo {author}
  {\bibfnamefont {G.}~\bibnamefont {Peng}},\ }\bibfield  {title} {\bibinfo
  {title} {Production of proton-rich nuclei in the vicinity of $^{100}${S}n via
  multinucleon transfer reactions},\ }\href
  {https://doi.org/https://doi.org/10.1016/j.physletb.2022.136886} {\bibfield
  {journal} {\bibinfo  {journal} {Phys. Lett. B}\ }\textbf {\bibinfo {volume}
  {825}},\ \bibinfo {pages} {136886} (\bibinfo {year} {2022})}\BibitemShut
  {NoStop}%
\bibitem [{\citenamefont {Chabanat}\ \emph {et~al.}(1998)\citenamefont
  {Chabanat}, \citenamefont {Bonche}, \citenamefont {Haensel}, \citenamefont
  {Meyer},\ and\ \citenamefont {Schaeffer}}]{Chabanat1998_NPA635-231}%
  \BibitemOpen
  \bibfield  {author} {\bibinfo {author} {\bibfnamefont {E.}~\bibnamefont
  {Chabanat}}, \bibinfo {author} {\bibfnamefont {P.}~\bibnamefont {Bonche}},
  \bibinfo {author} {\bibfnamefont {P.}~\bibnamefont {Haensel}}, \bibinfo
  {author} {\bibfnamefont {J.}~\bibnamefont {Meyer}},\ and\ \bibinfo {author}
  {\bibfnamefont {R.}~\bibnamefont {Schaeffer}},\ }\bibfield  {title} {\bibinfo
  {title} {A {S}kyrme parametrization from subnuclear to neutron star densities
  part {I}{I}. {N}uclei far from stabilities},\ }\href
  {https://doi.org/10.1016/S0375-9474(98)00180-8} {\bibfield  {journal}
  {\bibinfo  {journal} {Nucl. Phys. A}\ }\textbf {\bibinfo {volume} {635}},\
  \bibinfo {pages} {231} (\bibinfo {year} {1998})}\BibitemShut {NoStop}%
\bibitem [{\citenamefont {Sun}\ \emph {et~al.}(2022)\citenamefont {Sun},
  \citenamefont {Guo},\ and\ \citenamefont {Umar}}]{Sun2022_PRC105-034601}%
  \BibitemOpen
  \bibfield  {author} {\bibinfo {author} {\bibfnamefont {X.-X.}\ \bibnamefont
  {Sun}}, \bibinfo {author} {\bibfnamefont {L.}~\bibnamefont {Guo}},\ and\
  \bibinfo {author} {\bibfnamefont {A.~S.}\ \bibnamefont {Umar}},\ }\bibfield
  {title} {\bibinfo {title} {Microscopic study of the fusion reactions
  $^{40,48}\mathrm{Ca}+^{78}\mathrm{Ni}$ and the effect of the tensor force},\
  }\href {https://doi.org/10.1103/PhysRevC.105.034601} {\bibfield  {journal}
  {\bibinfo  {journal} {Phys. Rev. C}\ }\textbf {\bibinfo {volume} {105}},\
  \bibinfo {pages} {034601} (\bibinfo {year} {2022})}\BibitemShut {NoStop}%
\bibitem [{\citenamefont {Washiyama}(2015)}]{Washiyama2015_PRC91-064607}%
  \BibitemOpen
  \bibfield  {author} {\bibinfo {author} {\bibfnamefont {K.}~\bibnamefont
  {Washiyama}},\ }\bibfield  {title} {\bibinfo {title} {Microscopic analysis of
  fusion hindrance in heavy nuclear systems},\ }\href
  {https://doi.org/10.1103/PhysRevC.91.064607} {\bibfield  {journal} {\bibinfo
  {journal} {Phys. Rev. C}\ }\textbf {\bibinfo {volume} {91}},\ \bibinfo
  {pages} {064607} (\bibinfo {year} {2015})}\BibitemShut {NoStop}%
\bibitem [{\citenamefont {Bock}\ \emph {et~al.}(1982)\citenamefont {Bock},
  \citenamefont {Chu}, \citenamefont {Dakowski}, \citenamefont {Gobbi},
  \citenamefont {Grosse}, \citenamefont {Olmi}, \citenamefont {Sann},
  \citenamefont {Schwalm}, \citenamefont {Lynen}, \citenamefont
  {M{\ifmmode\ddot{u}\else\"{u}\fi}ller}, \citenamefont {Bj{\o}rnholm},
  \citenamefont {Esbensen}, \citenamefont
  {W{\ifmmode\ddot{o}\else\"{o}\fi}lfli},\ and\ \citenamefont
  {Morenzoni}}]{Bock1982_NPA388-334}%
  \BibitemOpen
  \bibfield  {author} {\bibinfo {author} {\bibfnamefont {R.}~\bibnamefont
  {Bock}}, \bibinfo {author} {\bibfnamefont {Y.~T.}\ \bibnamefont {Chu}},
  \bibinfo {author} {\bibfnamefont {M.}~\bibnamefont {Dakowski}}, \bibinfo
  {author} {\bibfnamefont {A.}~\bibnamefont {Gobbi}}, \bibinfo {author}
  {\bibfnamefont {E.}~\bibnamefont {Grosse}}, \bibinfo {author} {\bibfnamefont
  {A.}~\bibnamefont {Olmi}}, \bibinfo {author} {\bibfnamefont {H.}~\bibnamefont
  {Sann}}, \bibinfo {author} {\bibfnamefont {D.}~\bibnamefont {Schwalm}},
  \bibinfo {author} {\bibfnamefont {U.}~\bibnamefont {Lynen}}, \bibinfo
  {author} {\bibfnamefont {W.}~\bibnamefont
  {M{\ifmmode\ddot{u}\else\"{u}\fi}ller}}, \bibinfo {author} {\bibfnamefont
  {S.}~\bibnamefont {Bj{\o}rnholm}}, \bibinfo {author} {\bibfnamefont
  {H.}~\bibnamefont {Esbensen}}, \bibinfo {author} {\bibfnamefont
  {W.}~\bibnamefont {W{\ifmmode\ddot{o}\else\"{o}\fi}lfli}},\ and\ \bibinfo
  {author} {\bibfnamefont {E.}~\bibnamefont {Morenzoni}},\ }\bibfield  {title}
  {\bibinfo {title} {Dynamics of the fusion process},\ }\href
  {https://doi.org/10.1016/0375-9474(82)90420-1} {\bibfield  {journal}
  {\bibinfo  {journal} {Nucl. Phys. A}\ }\textbf {\bibinfo {volume} {388}},\
  \bibinfo {pages} {334} (\bibinfo {year} {1982})}\BibitemShut {NoStop}%
\bibitem [{\citenamefont {Newton}\ \emph
  {et~al.}(2004{\natexlab{a}})\citenamefont {Newton}, \citenamefont {Butt},
  \citenamefont {Dasgupta}, \citenamefont {Hinde}, \citenamefont {Gontchar},
  \citenamefont {Morton},\ and\ \citenamefont
  {Hagino}}]{Newton2004_PLB586-219}%
  \BibitemOpen
  \bibfield  {author} {\bibinfo {author} {\bibfnamefont {J.~O.}\ \bibnamefont
  {Newton}}, \bibinfo {author} {\bibfnamefont {R.~D.}\ \bibnamefont {Butt}},
  \bibinfo {author} {\bibfnamefont {M.}~\bibnamefont {Dasgupta}}, \bibinfo
  {author} {\bibfnamefont {D.~J.}\ \bibnamefont {Hinde}}, \bibinfo {author}
  {\bibfnamefont {I.~I.}\ \bibnamefont {Gontchar}}, \bibinfo {author}
  {\bibfnamefont {C.~R.}\ \bibnamefont {Morton}},\ and\ \bibinfo {author}
  {\bibfnamefont {K.}~\bibnamefont {Hagino}},\ }\bibfield  {title} {\bibinfo
  {title} {{Systematics of precise nuclear fusion cross sections: the need for
  a new dynamical treatment of fusion?}},\ }\href
  {https://doi.org/10.1016/j.physletb.2004.02.052} {\bibfield  {journal}
  {\bibinfo  {journal} {Phys. Lett. B}\ }\textbf {\bibinfo {volume} {586}},\
  \bibinfo {pages} {219} (\bibinfo {year} {2004}{\natexlab{a}})}\BibitemShut
  {NoStop}%
\bibitem [{\citenamefont {Newton}\ \emph
  {et~al.}(2004{\natexlab{b}})\citenamefont {Newton}, \citenamefont {Butt},
  \citenamefont {Dasgupta}, \citenamefont {Hinde}, \citenamefont {Gontchar},
  \citenamefont {Morton},\ and\ \citenamefont
  {Hagino}}]{Newton2004_PRC70-024605}%
  \BibitemOpen
  \bibfield  {author} {\bibinfo {author} {\bibfnamefont {J.~O.}\ \bibnamefont
  {Newton}}, \bibinfo {author} {\bibfnamefont {R.~D.}\ \bibnamefont {Butt}},
  \bibinfo {author} {\bibfnamefont {M.}~\bibnamefont {Dasgupta}}, \bibinfo
  {author} {\bibfnamefont {D.~J.}\ \bibnamefont {Hinde}}, \bibinfo {author}
  {\bibfnamefont {I.~I.}\ \bibnamefont {Gontchar}}, \bibinfo {author}
  {\bibfnamefont {C.~R.}\ \bibnamefont {Morton}},\ and\ \bibinfo {author}
  {\bibfnamefont {K.}~\bibnamefont {Hagino}},\ }\bibfield  {title} {\bibinfo
  {title} {Systematic failure of the woods-saxon nuclear potential to describe
  both fusion and elastic scattering: Possible need for a new dynamical
  approach to fusion},\ }\href {https://doi.org/10.1103/PhysRevC.70.024605}
  {\bibfield  {journal} {\bibinfo  {journal} {Phys. Rev. C}\ }\textbf {\bibinfo
  {volume} {70}},\ \bibinfo {pages} {024605} (\bibinfo {year}
  {2004}{\natexlab{b}})}\BibitemShut {NoStop}%
\bibitem [{\citenamefont {Cap}\ \emph {et~al.}(2021)\citenamefont {Cap},
  \citenamefont {Kowal},\ and\ \citenamefont
  {Siwek-Wilczy{\ifmmode\acute{n}\else\'{n}\fi}ska}}]{Cap2021_arXiv2107.00579}%
  \BibitemOpen
  \bibfield  {author} {\bibinfo {author} {\bibfnamefont {T.}~\bibnamefont
  {Cap}}, \bibinfo {author} {\bibfnamefont {M.}~\bibnamefont {Kowal}},\ and\
  \bibinfo {author} {\bibfnamefont {K.}~\bibnamefont
  {Siwek-Wilczy{\ifmmode\acute{n}\else\'{n}\fi}ska}},\ }\bibfield  {title}
  {\bibinfo {title} {Diffusion as a mechanism controlling the production of
  superheavy nuclei in cold fusion reactions},\ }\href
  {https://arxiv.org/abs/2107.00579v1} {\bibfield  {journal} {\bibinfo
  {journal} {arXiv}\ } (\bibinfo {year} {2021})},\ \Eprint
  {https://arxiv.org/abs/2107.00579} {2107.00579} \BibitemShut {NoStop}%
\bibitem [{\citenamefont {Cap}\ \emph {et~al.}(2014)\citenamefont {Cap},
  \citenamefont {Siwek-Wilczy{\ifmmode\acute{n}\else\'{n}\fi}ska},\ and\
  \citenamefont
  {Wilczy{\ifmmode\acute{n}\else\'{n}\fi}ski}}]{Cap2014_PLB736-478}%
  \BibitemOpen
  \bibfield  {author} {\bibinfo {author} {\bibfnamefont {T.}~\bibnamefont
  {Cap}}, \bibinfo {author} {\bibfnamefont {K.}~\bibnamefont
  {Siwek-Wilczy{\ifmmode\acute{n}\else\'{n}\fi}ska}},\ and\ \bibinfo {author}
  {\bibfnamefont {J.}~\bibnamefont
  {Wilczy{\ifmmode\acute{n}\else\'{n}\fi}ski}},\ }\bibfield  {title} {\bibinfo
  {title} {{No chance for synthesis of super-heavy nuclei in fusion of
  symmetric systems}},\ }\href {https://doi.org/10.1016/j.physletb.2014.07.062}
  {\bibfield  {journal} {\bibinfo  {journal} {Phys. Lett. B}\ }\textbf
  {\bibinfo {volume} {736}},\ \bibinfo {pages} {478} (\bibinfo {year}
  {2014})}\BibitemShut {NoStop}%
\bibitem [{\citenamefont {Jachimowicz}\ \emph {et~al.}(2021)\citenamefont
  {Jachimowicz}, \citenamefont {Kowal},\ and\ \citenamefont
  {Skalski}}]{Jachimowicz2021_ADNDT138-101393}%
  \BibitemOpen
  \bibfield  {author} {\bibinfo {author} {\bibfnamefont {P.}~\bibnamefont
  {Jachimowicz}}, \bibinfo {author} {\bibfnamefont {M.}~\bibnamefont {Kowal}},\
  and\ \bibinfo {author} {\bibfnamefont {J.}~\bibnamefont {Skalski}},\
  }\bibfield  {title} {\bibinfo {title} {Properties of heaviest nuclei with
  $98\leq{Z}\leq 126$ and $134\leq {N}\leq 192$},\ }\href
  {https://doi.org/10.1016/j.adt.2020.101393} {\bibfield  {journal} {\bibinfo
  {journal} {At. Data Nucl. Data Tables}\ }\textbf {\bibinfo {volume} {138}},\
  \bibinfo {pages} {101393} (\bibinfo {year} {2021})}\BibitemShut {NoStop}%
\end{thebibliography}

%

\end{document}